\documentclass[iop,twocolappendix,revtex4]{emulateapj}

\usepackage{color}
\usepackage{enumerate}
\usepackage{amsmath}
\usepackage{url}
\usepackage[colorlinks=true,linkcolor=black,citecolor=blue,urlcolor=blue,anchorcolor=black]{hyperref}

\newcommand{\myemail}{Ehsan.Moravveji@ster.kuleuven.be}

\shorttitle{Asteroseismic Modelling of KIC\,7760680}
\shortauthors{E. Moravveji et al. (2016)}

\begin{document}

\title{Sub-inertial gravity modes in the B8V star KIC\,7760680 
reveal moderate core overshooting and low vertical diffusive mixing}

\author{Ehsan Moravveji\altaffilmark{1}}
\affil{$^1$Institute of Astronomy, KU\,Leuven, Celestijnenlaan 200D, 
3001 Leuven, Belgium}
\altaffiltext{1}{\noindent Marie Curie post-doctoral fellow.}
\email{\myemail}
\and
\author{Richard H. D. Townsend\altaffilmark{2}}
\affil{$^2$Department of Astronomy, University of Wisconsin-Madison, Madison, WI 53706, USA}
\and
\author{Conny Aerts\altaffilmark{1,3}}
\affil{$^3$Department of Astrophysics, IMAPP, Radboud University Nijmegen, 
PO Box 9010, 6500 GL, Nijmegen, The Netherlands}
\and
\author{St\'ephane Mathis\altaffilmark{4}}
\affil{$^4$Laboratoire AIM Paris-Saclay, CEA/DSM - CNRS - Universit\'e Paris Diderot, 
IRFU/SAp Centre de Saclay, 91191, Gif-sur-Yvette Cedex, France}

%%%%%%%%%%%%%%%%%%%%%%%%%%%%%%%%%%%%%%%%%%%%%%%%%%%%%%%%%%%%%%%%%%%%%%%%%%%%%%%%%%%%%%%%%%%%%%%%%%%%

\begin{abstract}
KIC\,7760680 is so far the richest slowly pulsating B star, by exhibiting 36 consecutive dipole 
($\ell=1$) gravity (g-) modes.
The monotonically decreasing period spacing of the series, in addition to the local dips in the 
pattern confirm that KIC\,7760680 is a moderate rotator, with clear mode trapping in chemically 
inhomogeneous layers. 
We employ the traditional approximation of rotation to incorporate rotational effects on g-mode
frequencies.
Our detailed forward asteroseismic modelling of this g-mode series reveals that KIC\,7760680 is a 
moderately rotating B star with mass $\sim3.25$\,M$_\odot$.
By simultaneously matching the slope of the period spacing, and the number of modes in the 
observed frequency range, we deduce that the equatorial rotation frequency of KIC\,7760680 is 
0.4805\,day$^{-1}$, which is 26\% of its Roche break up frequency.
The relative deviation of the model frequencies and those observed is less than one percent.
We succeed to tightly constrain the exponentially-decaying convective core overshooting parameter to
$f_{\rm ov}\approx0.024\pm0.001$.
This means that convective core overshooting can coexist with moderate rotation.
Moreover, models with exponentially-decaying overshoot from the core outperform those with the 
classical step-function overshoot.
The best value for extra diffusive mixing in the radiatively stable envelope is confined to
$\log D_{\rm ext}\approx0.75\pm0.25$ (with $D_{\rm ext}$ in cm$^2$\,sec$^{-1}$), which is notably 
smaller than theoretical predictions.
%We deduce that KIC\,7760680 is a rigid body rotator, and the absence of differential rotation 
%has suppressed additional mixing in the radiative regions above the convective overshooting zone.
\end{abstract}

%%%%%%%%%%%%%%%%%%%%%%%%%%%%%%%%%%%%%%%%%%%%%%%%%%%%%%%%%%%%%%%%%%%%%%%%%%%%%%%%%%%%%%%%%%%%%%%%%%%%


%%%%%%%%%%%%%%%%%%%%%%%%%%%%%%%%%%%%%%%%%%%%%%%%%%%%%%%%%%%%%%%%%%%%%%%%%%%%%%%%%%%%%%%%%%%%%%%%%%%%

\section{Introduction}\label{s-intro}
Late- to mid-type B stars have masses between $\sim3$ to $7$\,M$_\odot$, and are pulsationally 
unstable against low-degree high-order $|n_{\rm pg}|\gtrsim10$ g-modes, with periods ranging
from $\sim0.5$ to $\sim3$ days.
They are classified as slowly pulsating B (SPB) stars
\citep{waelkens-1991-01,waelkens-1998-01,de-cat-2002-01,aerts-2010-book}.
Non-adiabatic heat exchange around the iron opacity bump at $\log T\approx5.2$\,K -- known
as the classical $\kappa-$mechanism -- is responsible for their mode excitation 
\citep{gautschy-1993-01,dziembowski-1993-02}.
Together with $\gamma$\,Dor stars, they are the richest main-sequence g-mode pulsators with 
numerous excited modes \citep[e.g. Fig.\,2e in][]{moravveji-2016-01}.
Chemically inhomogeneous regions above the receding convective cores reside inside this propagation
cavity, and allow for partial or complete g-mode trapping, providing a unique physical 
diagnostic of the chemical mixing and thermal stratification in the deep stellar interior 
\citep{miglio-2008-01,cunha-2015-01}.
In addition, g-modes can have sizeable amplitudes inside the narrow overshooting region 
-- between the fully-mixed convective core and the $\mu$-gradient layer, allowing to 
constrain the extent and physical properties of the overshooting layer 
\citep{dziembowski-1991-01, moravveji-2015-01,moravveji-2015-02}.
In such circumstances, g-modes can resolve and probe the physical conditions of the overshooting
layer.

Moderate to rapid rotation is a fairly established property of (single and binary) SPB stars
\citep{huang-2010-01}, and in general massive stars \citep{dufton-2013-01,ramirez-2013-01,
ramirez-2015-01}.
As soon as rotation kicks in, the centrifugal force deforms the star, and large-scale advection sets in.
Then, the thermal and structural equilibrium of stars undergo a readjustment, in order to conserve mass, 
energy, linear and angular momentum \citep{kippenhahn-1970-01,endal-1976-01,endal-1978-01,zahn-1992-01,
maeder-1998-01,mathis-2004-02,maeder-2009-book,espinosa-lara-2013-01}.
As a consequence of this, a handful of (advective and diffusive) mixing processes are triggered,
smoothing chemical inhomogeneities in radiative envelopes and transferring angular momentum between the 
stellar core and envelope \citep[see e.g.][]{endal-1978-01,heger-2000-01,heger-2005-01,maeder-2009-book}.
This poorly-known aspect of the theory of stellar evolution deserves a profound observational calibration.
From observational and theoretical standpoints, however, there is a long way ahead to test and 
grasp all proposed mixing mechanisms, and the possible interplay between them
\citep{heger-2000-01,maeder-2013-01}.
With asteroseismology of rotating and heat-driven pulsating stars, such as SPBs and $\gamma$\,Dor stars, 
we can quantitatively address several uncertain aspects of massive star evolution, and deep internal 
structure.
The very slowly rotating pulsating {\it Kepler} B8V star \objectname{KIC\,10526294} 
\citep[][hereafter Star\,\textsc{I}]{papics-2014-01} 
offered the first opportunity of seismic modelling of this type of pulsators.
In \cite{moravveji-2015-01}, we succeeded to place tight asteroseismic constraints on core 
overshooting, and extra diffusive mixing in the envelope of this star.
That was followed by the derivation of its internal differential rotation profile by 
\cite{triana-2015-01}, who inferred that the envelope of Star\,\textsc{I} rotates in the opposite
direction with respect to its core.

Pulsation instability among late B-type stars is a known phenomenon from ground-based photometric and
spectroscopic observations \citep{de-cat-2002-01}, 
in addition to CoRoT and {\it Kepler} space photometry \citep{papics-2011-01,papics-2012-01,balona-2011-01}. 
The impact of rotation on pulsation modes is profound, and is thoroughly explained in the literature. 
For instance, \cite{unno-1989-book} and \cite{townsend-2003-02} explain different classes of 
heat-driven inertial pulsation  modes that arise in rotating stars, like Rossby (r-) modes, 
Kelvin modes, and Yanai modes, because of the action of the Coriolis acceleration.
The centrifugal deformation of the star affects the low-density outer envelope more significantly than
the high-density core;
thus, p-modes are more influenced by the centrifugal force, while g-modes are influenced by the Coriolis force
\citep{dintrans-2000-01,reese-2006-01,ballot-2010-01}.
In addition to heat-driven modes destabilised by rotation, stochastic excitation of gravito-inertial 
waves was predicted by \cite{mathis-2014-01} and \cite{rogers-2013-01}, and observed by 
\cite{neiner-2012-01} in the CoRoT B0IVe target \objectname{HD\,51452}.
The feedback from low-frequency g-modes and r-modes in SPB stars is efficient transport of angular momentum 
\citep{rogers-2015-01,lee-2014-01,lee-2016-01}. 
Nonradial pulsations are even proposed to cause the Be phenomena by the energy leakage of 
low-frequency prograde g-modes \citep{shibahashi-2013-01}.
This was already observed in B0.5IVe CoRoT target \objectname{HD\,49330} \citep{huat-2009-01}.
Thus, rotation interacts profoundly with stellar structure, evolution and pulsation.

The pulsation description of rotating stars is at least a two-dimensional problem 
\citep[e.g.][]{prat-2016-01}.
However, it is possible to reduce this dimensionality into two separate one-dimensional problems for 
the radial \citep{lee-1986-01} and angular dependence \citep{lee-1997-01,townsend-2003-03} of 
eigensolutions.
That is achieved by ignoring the centrifugal deformation, and neglecting the horizontal 
component of the rotation vector in the momentum equation, when stratification dominates
rotation \citep[see e.g. the detailed discussion in][]{mathis-2008-01}.
This is historically known as the Traditional Approximation of Rotation \citep[TAR,][]{eckart-1960-book}.
Within the TAR framework, \cite{townsend-2005-01,townsend-2005-02} and \cite{aprilia-2011-01} have shown
that the combination of buoyancy and Coriolis forces, when coupled with the $\kappa$-mechanism due to
the iron-bump, provide a sufficient restoring force for driving high-order prograde $m=+1$
g-modes in SPB and $\gamma$\,Dor stars \citep{savonije-2005-01,bouabid-2013-01}.
In addition, these modes are predicted to exhibit significant photometric light variability,
and become observable \citep{townsend-2003-02,savonije-2013-01}.

The theoretical basis and observational facilities are now in place to exploit the wealth of
information contained in the frequency spectrum of rotating pulsating B stars.
The subject of this paper is the modelling and initial interpretation of a moderately rotating and 
slowly pulsating B8V star KIC\,7760680 that was recently discovered by \cite{papics-2015-01}.
We employ the identified series of dipole prograde g-modes in this star to address the
following basic questions regarding the internal structure and global evolution of massive stars:
(a) how to constrain the unknown rotation frequency of a rotating SPB star?
(b) does rotation suppress core overshooting?
(c) what are the combined and simultaneous effects of core overshoot mixing 
    and additional mixing in the radiatively stable envelopes of B stars?
(d) does the efficiency of the overshooting mixing decline radially from the fully mixed 
    convective core, or does it stay strongly efficient over a fraction of scale heights away from 
    the core boundary?    
Here, we provide answers to this set of questions, by a forward seismic modelling of our target star.

In Sect.\,\ref{s-obs} we introduce the seismic observables of KIC\,7760680 that we exploit, 
and justify using traditional approximation when modelling high-order g-modes.
The treatment of overshooting and extra diffusive mixing in the radiative envelope is the subject of 
Sect.\,\ref{s-mix}, followed by introducing the input physics of four asteroseismic grids of 
non-rotating models in Sect.\,\ref{s-mesa}.
The evolutionary models are computed with the one dimensional MESA stellar structure and 
evolution code \citep[][version\,7678]{paxton-2011-01,paxton-2013-01,paxton-2015-01}.
The TAR and a comparison with first-order frequency perturbation are presented in 
Sect.\,\ref{s-tar-ledoux}.
Our asteroseismic computations are performed with the GYRE (version\,4.2) linear nonradial 
adiabatic/nonadiabatic one dimensional pulsation code \citep{townsend-2013-01} that incorporates TAR.
We introduce a simple and robust scheme to optimise the unknown rotation frequency of the star
in Sect.\,\ref{s-optim-f-rot}, and statistically constrain the most likely rotation rate of the 
target.
The meric function that we use for model selection is discussed in Sect.\,\ref{s-chisq}.
In Sect.\,\ref{s-results}, we proceed to choose the best asteroseismic model that  
reproduces the observed slanted period spacing pattern, and elaborate on mode stability properties, 
and their efficient trapping in the overshooting region.
In Sect.\,\ref{s-end}, we summarise our findings, and discuss the missing input physics from 
current state-of-the-art one-dimensional evolutionary models, which need to be incorporated 
in the (near) future.

%%%%%%%%%%%%%%%%%%%%%%%%%%%%%%%%%%%%%%%%%%%%%%%%%%%%%%%%%%%%%%%%%%%%%%%%%%%%%%%%%%%%%%%%%%%%%%%%%%%%

\section{Asteroseismic Observables of KIC\,7760680}\label{s-obs}

% Figure %
\begin{figure}[t!]
\includegraphics[width=\columnwidth]{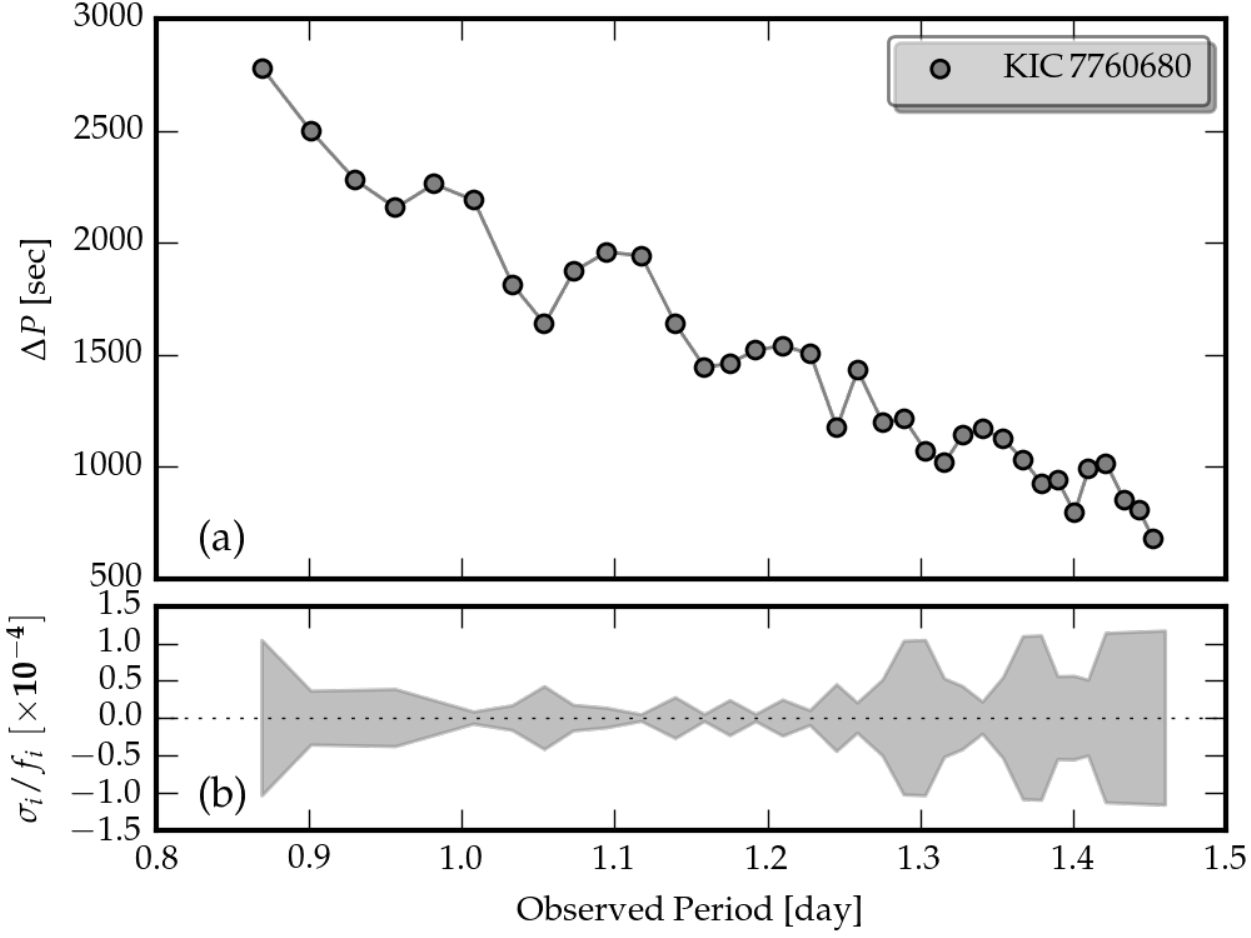} 
\caption{(a) The observed period spacing of KIC\,7760680, consisting of 36 dipole prograde g-modes.
For the list of observed modes refer to Table\,1 in \cite{papics-2015-01}.
The uncertainties are smaller than the plotting symbols.
(b) The observed relative frequency uncertainty $\sigma_i/f_i^{\rm(obs)}$.
Note that the ordinate is rescaled by a factor $10^{-4}$ for better visibility.
The modes in the middle of the series have the highest precision. 
\label{f-obs-dP}}
\end{figure}
% Figure %

\objectname{KIC\,7760680} (B8\,V) was observed by the nominal {\it Kepler} satellite for nearly four years.
\cite{papics-2015-01} carried out a thorough observational analysis of this target;
below, we summarise their findings relevant to our modelling. 
The spectroscopic properties of KIC\,7760680 from HERMES \citep{raskin-2011-01} high-resolution 
spectra are $T_{\rm eff}=11650\pm210$\,K, $\log g=3.97\pm0.08$\,dex, $[M/H]=0.14\pm0.09$, and 
$v\sin i=62\pm5$ km\,sec$^{-1}$.
The inferred $T_{\rm eff}$ and $\log g$ places KIC\,7760680 at the low-mass end of the SPB 
instability strip \citep{pamyatnykh-1999-01,moravveji-2016-01}.

\cite{papics-2015-01} identified a series of $\mathcal{N}=36$ low-frequency modes with periods 
between $P_{1}\pm\sigma_{1}=0.86930\pm0.00002$ and $P_{36}\pm\sigma_{36}=1.46046\pm0.00004$ 
days (their Table\,1).
This marks KIC\,7760680 as the richest SPB, discovered so far.
Fig.\,\ref{f-obs-dP}a shows the observed period spacing $\Delta P$, and Fig.\,\ref{f-obs-dP}b 
shows the relative frequency uncertainty $\sigma_i/f_i^{\rm(obs)}$ around each mode.
The striking feature of this series is the contiguous period spacing pattern with a negative 
slope.
In a non-rotating star, the asymptotic period spacing is 
$\Delta P_\ell^{\rm (asy)}=2\pi^2\left(\sqrt{\ell(\ell+1)}\int_{r_0}^{r_1}N/r\, dr\right)^{-1}$,
with $r_0$ and $r_1$ the inner and outer turning points of the mode propagation cavity, and the
Brunt-V\"ais\"al\"a frequency $N$ \citep{tassoul-1980-01}.
For a star with rotation frequency $f_{\rm rot}$, and pulsation g-mode frequency $f_i^{\rm(co)}$ 
in the co-rotating frame, the spin parameter is defined as $s_i=2f_{\rm rot}/f_i^{\rm(co)}$, 
and the period spacing $\Delta P_{\ell,m}^{\rm(co)}(s)$ is
\begin{equation}\label{e-dP}
\Delta P_{\tilde{\ell},m}^{\rm(co)}(s) \simeq \frac{2\pi^2}{\sqrt{\lambda_{\tilde{\ell},m,s(n+1)}} 
\,\int_{r_0}^{r_1}\frac{N}{r} dr \, \left(1+\frac{1}{2} \frac{d\ln\lambda_{\tilde{\ell},m,s(n)}}{d\ln s}\right)},
\end{equation}
where $n,\tilde{\ell},m$ are mode wavenumbers, and $\lambda_{\tilde{\ell},m,s}$ is the eigenvalue of the Laplace
Tidal equation \citep{townsend-2003-03,ballot-2012-01,bouabid-2013-01}.
For $f_{\rm rot}=0$, $\lambda_{\tilde{\ell},m,s}$ reduces to $\ell(\ell+1)$.
Based on Eq.\,(\ref{e-dP}), period spacing depends sensitively on the thermal and chemical 
stratification through the Brunt-V\"ais\"al\"a frequency $N$, in addition to the star's 
rotation and pulsation frequencies through the spin parameter $s$.
Thus, it provides a powerful asteroseismic diagnostic for constraining star's internal structure, 
in addition to its rotation frequency.
Here, we attempt to model the observed pulsation freqeuncies, and reproduce the period spacing in 
Fig.\,\ref{f-obs-dP}a.

By a visual inspection of the observed period spacing pattern in Fig.\,\ref{f-obs-dP}a, 
two important inferences follow:
(a) the negative moderate slope unravels the fact that KIC\,7760680 is a moderate rotator.
    A comparison with Fig.\,4 in \cite{bouabid-2013-01} and Figs.\,4 \& 5 in \cite{vanreeth-2015-02} 
    manifests that this series belongs to prograde modes, which are also theoretically predicted 
    to be unstable in rotating SPB stars \citep{townsend-2005-01,townsend-2005-02,aprilia-2011-01}. 
(b) there are clear deviations from the (tilted) asymptotic period spacing, which manifest 
    themselves as local dips.
    The reason behind this is the presence of an additional bump in the Brunt-V\"ais\"al\"a 
    frequency, associated with the $\mu$-gradient zone above the core \citep{miglio-2008-01}, 
    giving rise to efficient mode trapping in this region (discussed in Sect.\,\ref{ss-trap}).
    This inference was shown earlier after a detailed asteroseismic modelling of Star\,\textsc{I} 
    \citep[Fig.\,2 in ][]{moravveji-2015-02}.
    This proves that the mixing in the radiative envelope is not strong enough to chemically 
    homogenise the radiative zone.
    The presence of local dips in the observed period spacing puts an upper limit on the 
    effective amount of chemical mixing in the radiative envelope of this rotating SPB.

The list of 36 dipole g-modes of KIC\,7760680 were determined following the methodology discussed in 
detail in \citet{degroote-2009-01}, which is based on the theory of time series analysis of 
correlated data \citep{schwarzenberg-czerny-1991-01}. 
Practically, we take the formal errors of the non-linear least-squares fit to the light curve, 
and correct them for the signal-to-noise ratio, sampling, and correlated nature of the data. 
Based on this procedure explained in \citeauthor{degroote-2009-01}, a correction factor of $Q=$4.0 
(P.\,I.\,P\'apics, private communication) is applied to the formal errors listed in Table\,1 of 
\cite{papics-2015-01}.

For heat-driven pulsators, unambiguous mode identification from ({\it Kepler}) white-light photometry 
is only possible if one detects (almost) equally-spaced frequency splittings around isolated peaks.
Examples of such can be found in \cite{papics-2014-01}, \cite{kurtz-2014-01} and \cite{saio-2015-01}.
While this was feasible for Star\,\textsc{I}, \cite{papics-2015-01} could not discern any frequency
splitting for KIC\,7760680, due to the very high density of peaks in the narrow g-mode frequency 
domain.
Thus, one cannot assume any harmonic degree $\ell$ and azimuthal order $m$ for the detected series.
To tackle this, we computed few evolutionary tracks that pass through the 1$\sigma$ position of the 
star on the Kiel diagram, and chose a model that closely reproduced the observed period spacing, 
after including rigid rotation.
Then, we computed period spacing patterns for all possible combinations of $1\leq\ell\leq2$ and 
$|m|\leq\ell$ to explain the detected spacing. 
Appendix\,\ref{ap-em} and Fig.\,\ref{f-dP-all-l-m} present the results.
The only possible way to simultaneously reproduce the slope of the observed series, 
the number of observed modes inside the observed range $\mathcal{N}$, and the location of the input 
model on the Kiel diagram (inside the 1$\sigma$ spectroscopic box) is if the observed series be 
associated with dipole prograde $(\ell, \,m)=(1, \,+1)$ g-modes, which we adopt in what follows.

%%%%%%%%%%%%%%%%%%%%%%%%%%%%%%%%%%%%%%%%%%%%%%%%%%%%%%%%%%%%%%%%%%%%%%%%%%%%%%%%%%%%%%%%%%%%%%%%%%%%

\section{A Simplified Mixing Scheme}\label{s-mix}

Instead of exploiting non-exhaustive lists of proposed rotational and non-rotational mixing 
mechanisms \citep[e.g.][]{heger-2000-01,heger-2005-01,maeder-2013-01,mathis-2013-01}, we take a 
pragmatic approach and divide non-convective sources of mixing, and their corresponding coefficients 
into two distinct categories:
(a) overshooting from the convective core into radiative interior in a diffusive regime 
    $D_{\rm ov}$, and 
(b) an effective extra diffusive mixing $D_{\rm ext}$ from top of the overshoot layer up to the 
    surface. 
    Hereafter, $D_{\rm ext}$ is in cm$^2$\,sec$^{-1}$.
Fig.\,\ref{f-Ov-Dmix}a depicts our adopted mixing scheme. 
In the core, the convective mixing (blue) is computed from the Mixing Length Theory 
\citep[MLT,][]{bohm-vitense-1958-01,cox-1968-01};
there, the temperature gradient is almost adiabatic $\nabla\simeq\nabla_{\rm ad}$.
The overshoot region (grey) is installed at the outer boundary of the convective core. 
MLT does not apply in this region, and $D_{\rm ov}$ is instead calculated from an ad hoc prescription. 
In the present work, we consider two prescriptions offered by MESA:

% Figure %
\begin{figure*}[t!]
\begin{minipage}{0.48\textwidth}
\includegraphics[width=\columnwidth]{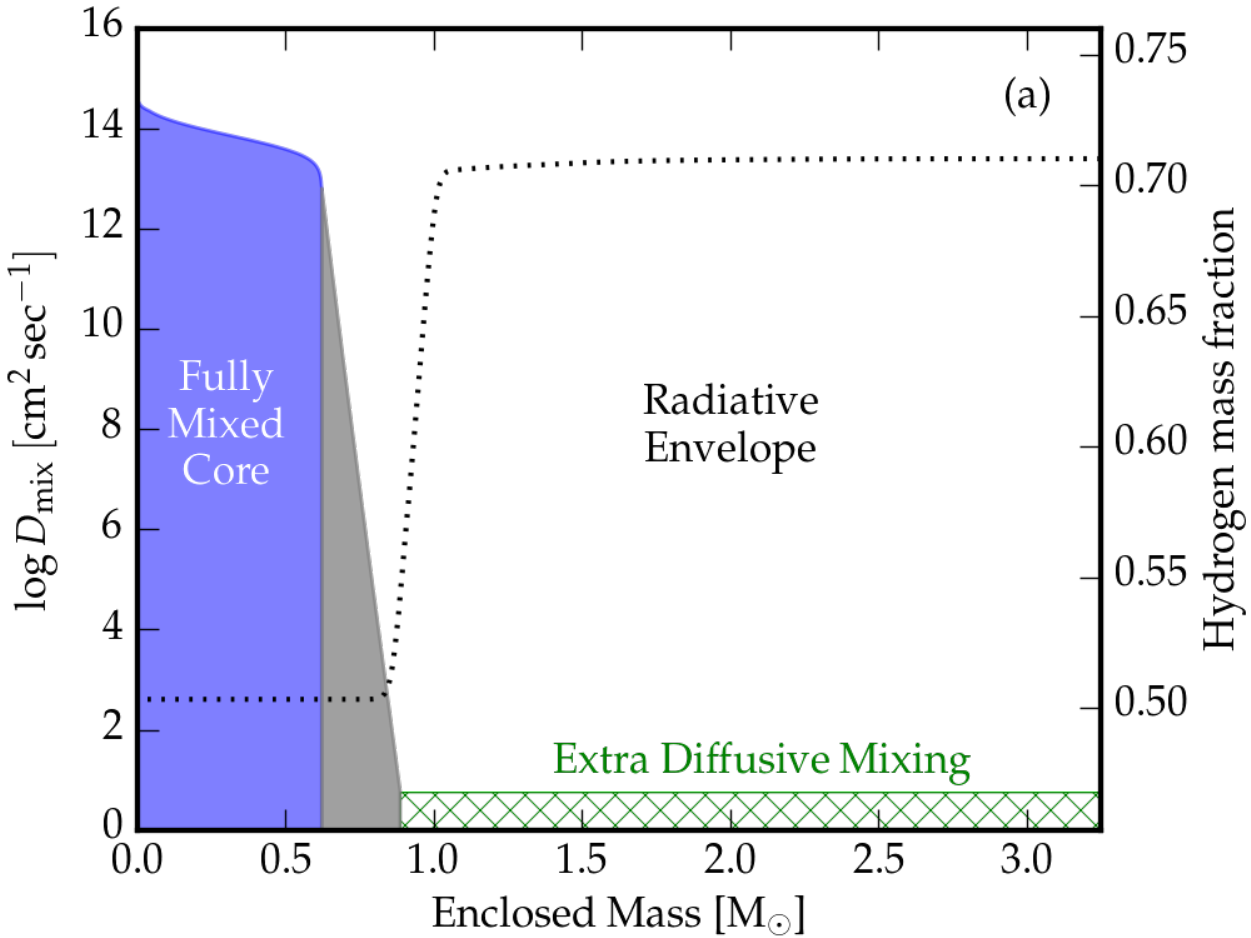}
\end{minipage}
\hspace{0.02\textwidth}
\begin{minipage}{0.48\textwidth}
\includegraphics[width=\columnwidth]{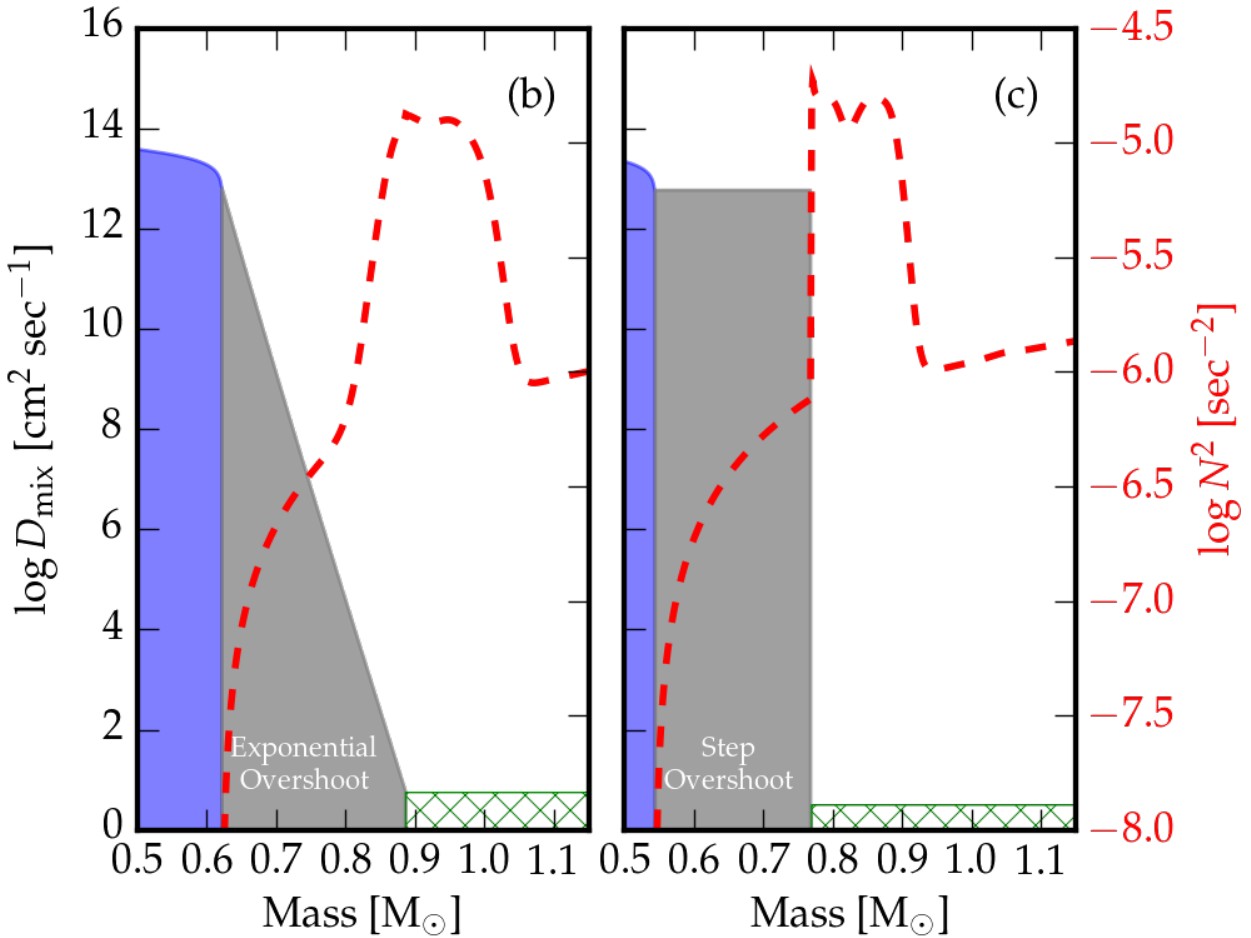} 
\end{minipage}
\caption{(a) The simplified mixing scheme in our evolutionary models, shown for the best asteroseismic 
model of KIC\,7760680 (discussed in Sect.\,\ref{ss-best} and Table\,\ref{t-chisq}).
The abscissa is the mass coordinate, and the ordinate is the logarithm of the mixing coefficient.
The convective and overshooting regions are shown in blue and grey, respectively.
Beyond the overshooting region, we include an additional diffusive mixing with varying strength, 
(green hatch).
The profile of hydrogen mass fraction (dotted lines) shows that the overshoot region is fully mixed.
(b) A zoom-in view around the overshoot region, for the best model with exponentially decaying prescription 
(Eq.\,\ref{e-fov}).
Grid A is built on this option.
(c) Similarly, for the best model with step-function overshooting.
Grid B is built on this option.
Notice the difference between the two Brunt-V\"ais\"al\"a profiles (red dashed line), and the extent
of the two mixing regions.
\label{f-Ov-Dmix}}
\end{figure*}
% Figure %

\begin{enumerate}
\renewcommand{\theenumi}{\Alph{enumi}}
\item Exponential overshoot \citep[after][]{freytag-1996-01,herwig-2000-01}, where the diffusion 
      coefficient for overshoot has a radial dependence
      \begin{equation}\label{e-fov}
         D_{\rm ov} = D_{\rm conv} \exp \left( - \frac{2 (r - r_{0})}{f_{\rm ov} H_{p}} \right) \qquad r_{0} \le r.
      \end{equation}
      Here, $r_{0} = r_{\rm cc} - f_{0} H_{p}$ is the radial coordinate of the lower boundary of the 
      overshoot region, which is situated at a depth $f_{0} H_{p}$ below the radius $r_{\rm cc}$ of 
      the convective core boundary; 
      $H_{p}$ is the pressure scale height, evaluated at $r_{\rm cc}$, and $D_{\rm conv}$ is the 
      convective mixing diffusion coefficient, evaluated from MLT at $r_{\rm 0}$. 
      The dimensionless parameters $f_{0}$ and $f_{\rm ov}$ allow tuning of the position and 
      exponential scale, respectively, of the overshoot region. 
      We fix $f_{0} = 10^{-3}$ throughout the present work, but allow $f_{\rm ov}$ to vary.
\item Step overshoot, where the diffusion coefficient for overshoot has a fixed value
      \begin{equation}\label{e-aov}
         D_{\rm ov} = D_{\rm conv} \qquad r_{0} \le r \le r_{0} + \alpha_{\rm ov} H_{p}.
      \end{equation}
      The interpretations of $r_{0}$, $H_{p}$ and $D_{\rm conv}$ are the same as before, 
      and again we adopt $f_{0} = 10^{-3}$ throughout; 
      but now the parameter $\alpha_{\rm ov}$ tunes the extent of the overshooting zone.
\end{enumerate}

In the overshoot region, MESA assumes $\nabla = \nabla_{\rm rad}$;
this differs from penetrative overshoot treatments \citep[e.g.][]{roxburgh-1965-01,maeder-1975-01,
zahn-1991-01,viallet-2015-01}, which are similar to the step prescription B (above) but assume 
$\nabla = \nabla_{\rm ad}$ over $d_p$ a penetration distance derived in \cite{zahn-1991-01}.
The time dependent turbulent convection model of \cite{zhang-2016-01} allows to set $\nabla$
between $\nabla_{\rm ad}$ and $\nabla_{\rm rad}$, but those models are not confronted with
observations, yet.

In Figs.\,\ref{f-Ov-Dmix}b and \ref{f-Ov-Dmix}c, we present the exponentially decaying
and step-function prescriptions, respectively.
The two best seismic models of our target (discussed later in Sect.\,\ref{ss-best}) are used
as input.
For the former, the transition from overshoot to extra mixing in the envelope is smooth.
On the other hand, the step-function overshoot mixing, as in Fig.\,\ref{f-Ov-Dmix}b implies 
constant mixing inside the overshoot zone, and suddenly drops by $\sim$13 to 7 orders of magnitude,
depending on what we adopt for $D_{\rm ext}$.
Here, the transition from overshoot to extra mixing is not smooth.
The resulting Brunt-V\"ais\"al\"a frequencies $N^2$ are shown in both panels with red 
dashed lines.
The difference between the two adopted overshoot prescriptions is that in the case of step-function 
overshoot, $N^2$ rises steeply at the top of the overshoot layer, whereas in the
exponentially decaying model, $N^2$ grows smoothly.
These two different $N^2$ profiles result in two different period spacing patterns, and 
allow discriminating them when modelling SPB stars.

Finally, a constant extra diffusive mixing (green hatch) is applied above the overshooting region 
across the remaining part of the radiative envelope (with $\nabla=\nabla_{\rm rad}$, see 
Fig.\,\ref{f-Ov-Dmix}).
The extra mixing can be associated with advecto-diffusive mixing due to rotation instability
\citep{heger-2000-01,maeder-2009-book}, mixing by an internal magnetic field 
\citep{heger-2005-01, mathis-2005-01}, semi-convective mixing \citep{langer-1985-01}, internal 
gravity waves \cite{talon-2005-01,pantillon-2007-01,rogers-2013-01,rogers-2015-01}, and other 
possible sources including their complex interaction \citep[e.g.][]{maeder-2013-01}.
Our proposed scheme is the least model-dependent approach to quantify the order-of-magnitude
of the non-convective (diffusive) mixing coefficients beyond the fully mixed core in B stars.

Because both $D_{\rm ov}$ and $D_{\rm ext}$ are unconstrained from first principles, we parametrize
them, and scan the parameter space to find the optimal values that explain the observed 
pulsation frequencies (or equivalently period spacing) of our target.
We recently developed this approach in \cite{moravveji-2015-01}, and carried out a detailed forward
modelling of nineteen dipole ($\ell=1$) g-modes in Star\,\textsc{I} (B8\,V, $v\sin i<18$\,km\,sec$^{-1}$).
For this specific star, we found the diffusive exponential overshooting prescription more favorable 
than the step-function prescription, and confined the free overshooting parameter to 
$f_{\rm ov}\approx0.017$.
Moreover, including extra diffusive mixing with coefficient of $\log D_{\rm ext}=1.75$ 
improved the quality of frequency fitting by a factor $\sim11$.
This was the first asteroseismic quantification of extra mixing in B stars.
Here, we apply the same methodology to KIC\,7760680, which rotates much faster than Star\,\textsc{I}.

%%%%%%%%%%%%%%%%%%%%%%%%%%%%%%%%%%%%%%%%%%%%%%%%%%%%%%%%%%%%%%%%%%%%%%%%%%%%%%%%%%%%%%%%%%%%%%%%%%%%

\section{Asteroseismic Models}\label{s-mesa}

We compute non-rotating non-magnetic stellar structure and evolution models with MESA 
with the mixing scheme described in Sect.\,\ref{s-mix}.
Each evolutionary track is specified with the following three parameters at zero-age main sequence:
the initial mass M$_{\rm ini}$, core overshooting free parameter $f_{\rm ov}$ (for exponential 
overshoot) or $\alpha_{\rm ov}$ (for step-function overshoot), and extra diffusive mixing beyond 
the overshoot region $D_{\rm ext}$.
All models assume the \cite{asplund-2009-01} metal mixture with the initial hydrogen mass fraction 
X$_{\rm ini}=0.71$ taken from the Galactic B-star standard of \cite{nieva-2012-01}.
We vary the initial metallicity Z$_{\rm ini}\in[0.014, 0.023]$;
the initial helium abundance is then fixed accordingly Y$_{\rm ini}$=1-X$_{\rm ini}$-Z$_{\rm ini}$.
Along every evolutionary track, we store an equilibrium model at every $\sim0.001$ drop in 
X$_{\rm c}$.
We terminate the evolution as soon as X$_{\rm c}$ drops below 10$^{-3}$.
The convective boundaries are specified using the Ledoux criterion.
For the surface boundary condition, we use ATLAS\,9 tables of \cite{castelli-2003-01} with surface
optical depth $\tau_{\rm s}=2/3$.
We include the line-driven mass loss prescription of \cite{vink-2001-01} with the efficiency factor
reduced by a factor 3 \citep{puls-2015-01}.

Recently, \cite{moravveji-2016-01} showed that a 75\% increase in Iron and Nickel monochromatic
opacities explains the position of two confirmed $\beta$\,Cep and eight confirmed hybrids on the 
Kiel diagram, which could not be explained before.
The increase resulted from the direct laboratory Iron opacity measurement of \cite{bailey-2015-01},
which was later confirmed by numerical simulations of \cite{nagayama-2016-01}.
We use this set of OP Iron- and Nickel-enhanced opacity tables, because it solves the pulsation
instability problem in massive stars, in agreement with previous predictions 
\citep{dziembowski-2008-01,salmon-2012-01}.
The recent OPAS opacity computation of \cite{mondet-2015-01} independently shows that the iron 
opacity is underestimated in default OP \citep{seaton-2005-01,badnell-2005-01} tables by 
$\sim40\%$ (their discussion in Sect.\,5).
Our computations are based on a 75\% Fe and Ni enhanced opacity tables.
The MESA inlists, and the opacity tables are freely available for download.
More information is provided in Appendix\,\ref{ap-files}.

% Table %
\begin{table}[t]
\caption{
Parameters of the two asteroseismic grids for KIC\,7760680.
We vary the initial mass M$_{\rm ini}$, exponential (or step-function)
overshoot $f_{\rm ov}$ (or $\alpha_{\rm ov}$), initial metallicity Z$_{\rm ini}$,
extra diffusive mixing $\log D_{\rm ext}$ (in cm$^2$ sec$^{-1}$), and core hydrogen 
mass fraction (X$_{\rm c}$).
$N$ is the total number of values for each parameter.
The number of degrees of freedom for all grids is $n=5$.
``Step" gives the minimum stepsize in the corresponding parameter.
} 
\label{t-grids}
\begin{center}
		\begin{tabular}{lllll}
			\hline
			Grid   & From & To     & Step & $N$ \\
			\hline
			{\bf A} \\
			M$_{\rm ini}$ [M$_\odot$] & 3.00  & 3.60  & 0.05 & 13 \\
			$f_{\rm ov}$  & 0.007 & 0.031 & $\geq$0.001 & 13 \\
			$Z_{\rm ini}$ & 0.014 & 0.023 & $\geq$0.001 & 8 \\
			$\log D_{\rm ext}$ & None & 5.0 & $\geq$0.25 & 13 \\ % logD=0 means no extra mixing
			X$_{\rm c}$        & 0.70 & 0.30 & $\geq$0.001 & $\sim$401 \\ 
            \hline
			{\bf B} \\    % alpha_ov
			M$_{\rm ini}$ [M$_\odot$] & 3.00  & 3.40  & 0.05 & 9 \\ % 
			$\alpha_{\rm ov}$ & 0.21 & 0.33 & $\geq$0.01 & 9 \\ % with an additional zoom with 0.29 and 0.31
			$Z_{\rm ini}$ & 0.014 & 0.023 & $\geq$0.001 & 4 \\
			$\log D_{\rm ext}$ & None & 1.50 & $\geq$0.25 & 4 \\
			X$_{\rm c}$        & 0.60 & 0.40 & $\geq$0.001 & $\sim$201 \\
			\hline
		\end{tabular}
\end{center}
\end{table}
% Table %

Based on the above setup, we compute two evolutionary grids, with the range and stepsize of 
parameters listed in Table\,\ref{t-grids}.
The procedure is to start from a coarse parameter space, and iteratively zoom around the best 
parameter ranges by decreasing the parameter stepsize.
The physical setup of both grids are identical, except for the choice of overshoot prescription.
The exponential prescription is employed in grid A, where we vary $f_{\rm ov}$ (see 
Fig.\,\ref{f-Ov-Dmix}b);
similarly, the step-function prescription is used in grid B, where we vary $\alpha_{\rm ov}$
 (see Fig.\,\ref{f-Ov-Dmix}c).
This allows to assess which of the two overshooting prescriptions is superior, in the sense of 
providing a better fit to the observed frequencies.
For Star\,\textsc{I}, we demonstrated that the exponential prescription outperformed the step-function
one \citep{moravveji-2015-01}.

Extra diffusive mixing is one of the grid parameters that we vary from $\log D_{\rm ext}=0.25$ to 5.0.
We also include models suppressing this;
these are presented in Table\,\ref{t-grids} and the forthcoming figures with ``None".
With this choice, we can examine if extra diffusive mixing is required in the envelope of SPB stars.
For Star\,\textsc{I}, the $\chi^2$ scores were reduced by a factor of more than 11 when we 
included extra mixing.
Here, we re-examine this for KIC\,7760680.

%%%%%%%%%%%%%%%%%%%%%%%%%%%%%%%%%%%%%%%%%%%%%%%%%%%%%%%%%%%%%%%%%%%%%%%%%%%%%%%%%%%%%%%%%%%%%%%%%%%%

\section{Traditional Approximation versus First-Order Perturbations}\label{s-tar-ledoux}

There are two possible approaches to incorporate the effect of rotation on pulsation frequencies based
on 1D stellar models:
one is through first-, second- and third-order perturbative methods
\citep{ledoux-1951-01,dziembowski-1992-01,soufi-1998-01}, and the other is through TAR.
To assess the validity of the first-order perturbative approach \citep{ledoux-1951-01} versus
the TAR, one should consider the spin parameter $s$.
\cite{ballot-2010-01} showed that for p-modes and low-order g-modes, it is still 
possible to use first-, second-, and/or third-order frequency corrections to reproduce the 
results from TAR, for low values of the spin parameter $s \ll 1$
\citep[refer to][for $\gamma$\,Dor stars]{bouabid-2013-01}.
As soon as $s_i\gtrsim1$, the frequency splittings from perturbative methods depart from 
their counterparts within the TAR framework, due to ignoring the impact of the Coriolis force
on pulsation frequencies.
This is the case for high-order g-modes -- even in slowly rotating stars -- due to their 
small frequency values.

\begin{figure}[t!]
\includegraphics[width=\columnwidth]{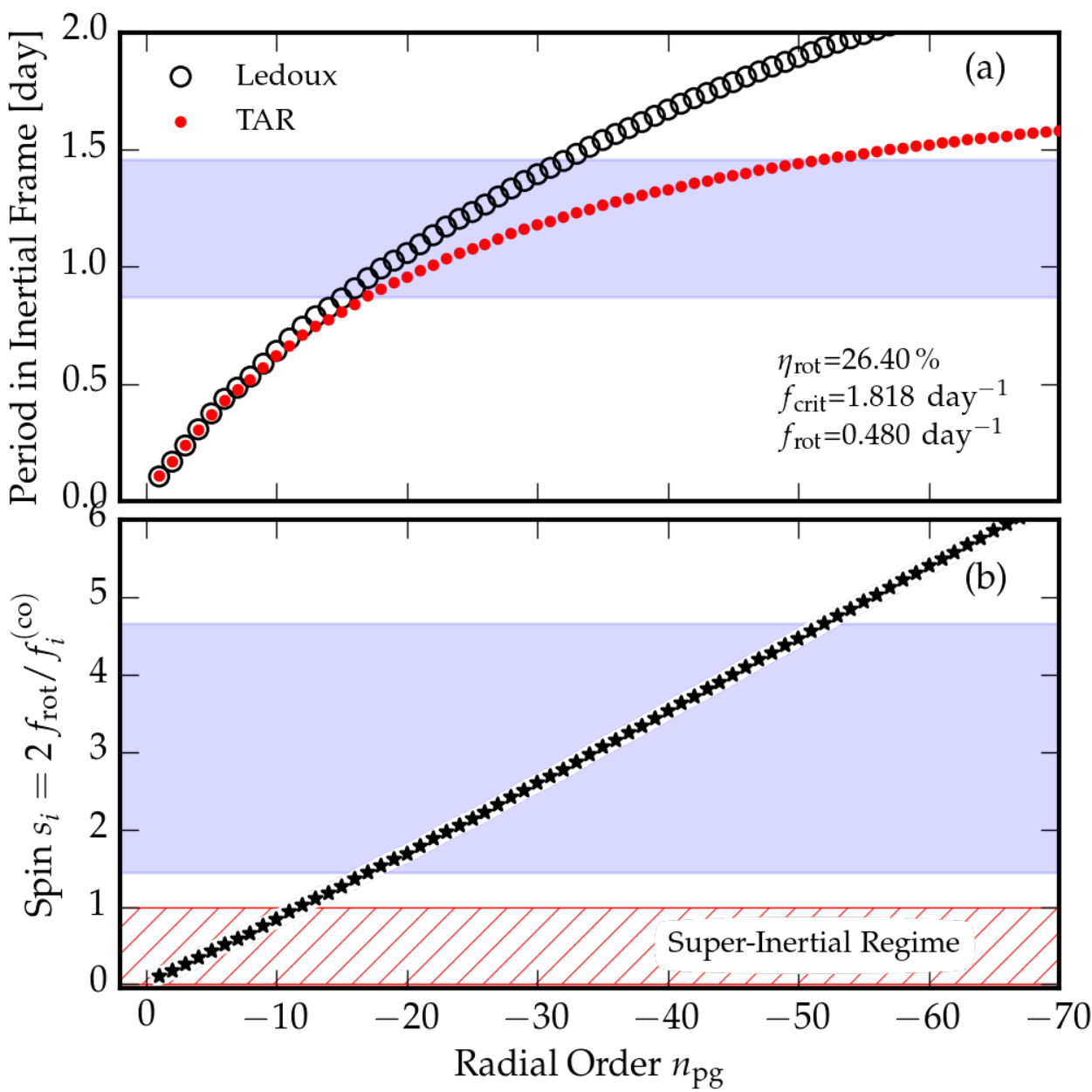} 
\caption{Top. First order perturbative periods (empty circles) versus those computed within TAR 
(red dots) in the inertial frame.
The blue band highlights the observed period range of KIC\,7760680.
The equilibrium structure of the model is adapted to the best asteroseismic model of 
KIC\,7760680 (Sect.\,\ref{s-results}), and is set to rotate rigidly at 26.4\% of the Roche critical 
frequency.
Bottom. The spin parameter for the corresponding mode frequencies in the co-rotating frame, $f_i^{\rm (co)}$.
The observed modes (in blue band) fall outside the super-inertial regime $s\leq1$ (red hatch),
and lie in the sub-inertial regime $s>1$.
\label{f-spin}}
\end{figure}

To demonstrate this, Fig.\,\ref{f-spin}a compares periods of dipole prograde modes from the first-order 
perturbative method (open circles) versus those from TAR (red dots), in the observer (inertial) 
reference frame.
The spin parameter for the corresponding modes in the co-rotating frame is shown in Fig.\,\ref{f-spin}b.
The input model corresponds to the best asteroseismic model of KIC\,7760680 (to be discussed later in 
Sect.\,\ref{s-results}), and is set to rotate rigidly at 26.4\% of the Roche break up frequency.
The range of the observed modes is shown by the blue band.
The difference between the two sets of periods is considerable, and the resulting period spacing 
-- which is tangent to each of these curves -- will significantly differ.
For sub-inertial waves, the wave dynamics is modified compared to TAR computations 
\citep{mathis-2008-01,mathis-2014-01};
however, \cite{ballot-2012-01} have shown that the resulting period spacing within TAR gives a correct
prediction up to high spin parameters.
Consequently, even for sub-inertial g-modes $s>1$, we decided to employ the TAR.

% Figure %

Even though the centrifugal force implies deviations from spherical symmetry, this effect becomes
important for stars rotating more than half critical \citep[Fig.\,1 in][]{aerts-2004-02}.
Using Eqs.\,1 and 2 in \cite{aerts-2004-02}, the polar radius of our target is smaller than its
equatorial radius by only $\sim1.6\%$, when the star is set to rotate at 26\% Roche critical frequency.
Moreover, the stellar surface of slower rotators can deform significantly due to the effect of 
centrifugal force, the deep interior -- where high-order g-modes propagate with larger amplitudes -- 
departs negligibly from spherical symmetry;
see  e.g. \cite{saio-2012-01} for a demonstration.
This means that modelling and studying the structure of the core overshooting layer in slow to moderate 
rotators with high-order g-modes, using one-dimensional stellar structure 
models coupled with one-dimensional oscillation theory under TAR, and ignoring the centrifugal
deformation is fully justified.

%%%%%%%%%%%%%%%%%%%%%%%%%%%%%%%%%%%%%%%%%%%%%%%%%%%%%%%%%%%%%%%%%%%%%%%%%%%%%%%%%%%%%%%%%%%%%%%%%%%%

\section{Opmtimizing Rotation Frequency}\label{s-optim-f-rot}

The lack of prior knowledge on the inclination angle of the rotation axis of our target precludes 
deducing the equatorial rotation frequency $f_{\rm rot}$ from the spectroscopic 
measurement of the projected rotation velocity $v\sin\,i=2\pi R_\star f_{\rm rot}\sin i=62\pm5$ 
km\,sec$^{-1}$, assuming a reasonable radius $R_\star$ from models.
Thus, $f_{\rm rot}$ is an additional unknown of KIC\,7760680.
Because MESA supports shellular rotation \citep{paxton-2013-01}, $f_{\rm rot}$ could be treated as 
another free parameter in our grids.
However, we choose not to do so, because, through their dependence on the spin parameter $s_i$, 
the g-mode frequencies are very sensitive to even a slight change in $f_{\rm rot}$ in the co-rotating 
and inertial frames.
As a demonstration, see Fig.\,2 in \cite{townsend-2005-01} or Fig1.\,1 and 2 in 
\cite{bouabid-2013-01}.
This would require an unreasonably broad and immensely resolved parameter survey for $f_{\rm rot}$, 
which is not computationally feasible.
Instead we take a pragmatic approach, and optimise $f_{\rm rot}$ for every input model. 
One can benefit from the observational fact that there are exactly $\mathcal{N}=36$ observed 
modes between $f_1$ and $f_{36}$ (allowing for a tolerance around them).
Thanks to the high sensitivity of $\mathcal{N}$ to $f_{\rm rot}$, we can tune the latter
until $\mathcal{N}=36$ is satisfied.

% Figure %
\begin{figure}[t!]
\includegraphics[width=\columnwidth]{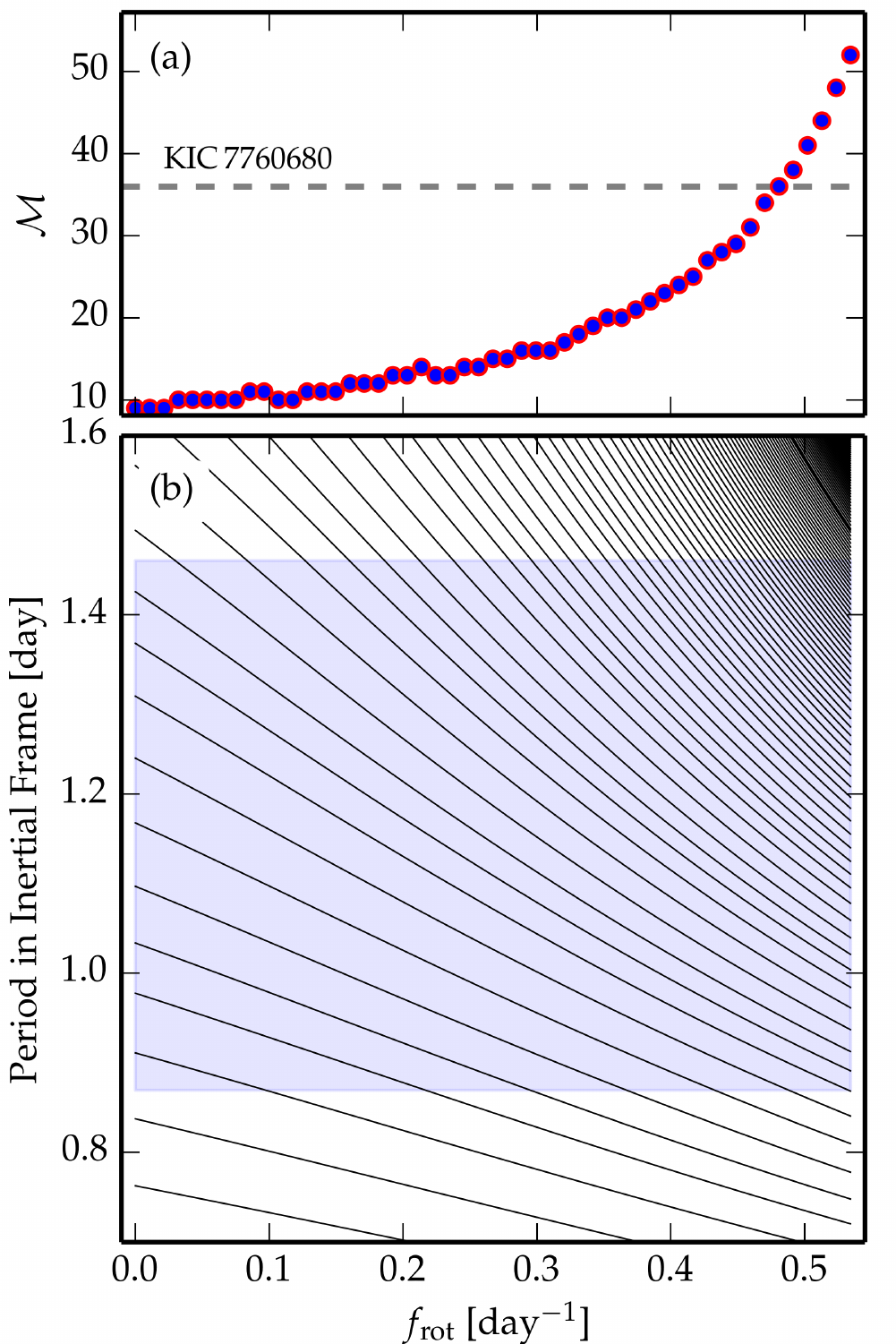} 
\caption{Top. The number of modes (circles) between the observed range (dashed line)
$\mathcal{N}$ versus rotation frequency $f_{\rm rot}$ in the inertial frame.
Bottom. Evolution of the mode periods in the inertial frame $P_i^{\rm(in)}$ versus $f_{\rm rot}$, 
for dipole prograde modes.
We use the best model (Sect.\,\ref{ss-best} and Table\,\ref{t-chisq}) as input.
The observed range $P_1\leq P_i^{\rm(in)}\leq P_{36}$ is highlighted in blue.
%The red dashed line shows $s=1$.
For higher $f_{\rm rot}$, a dense spectrum of high-order g-modes enters the observed range from 
the top, rapidly increasing $\mathcal{M}$.
%The black thick line is actually two close periods undergoing avoided crossing.
\label{f-optim-f-rot}}
\end{figure}
% Figure %

Fig.\,\ref{f-optim-f-rot}b shows the evolution of mode periods in the inertial frame versus
rotation frequency $f_{\rm rot}$.
The blue band highlights the observed range.
%The red dashed curve markes $s=1$, and separates the modes in the gravito-inertial regime
%$s<1$ (left), from those in the inertial regime $s>1$ (right).
Clearly, increasing the rotation frequency (and hence the spin parameter) progressively 
increases the number of modes inside the observed range.
At the same time few modes gradually leave the observed range. 
The net number of modes inside the observed range $\mathcal{M}$ is shown in 
Fig.\,\ref{f-optim-f-rot}a, showing the strong dependence of $\mathcal{M}$ on $f_{\rm rot}$.
The dashed horizontal line also shows the observed number of modes for KIC\,7760680, i.e. 
$\mathcal{N}=36$.
We define $d\mathcal{N}(f_{\rm rot})$ as an integer-valued function that captures the difference 
between the model and observed number of modes in the inertial frame within the observed range. 
\begin{equation}\label{e-dN}
d\mathcal{N}(f_{\rm rot}) = \mathcal{M} - \mathcal{N},
\end{equation}
Consequently, $d\mathcal{N}$ can be used as a discriminant to {\it optimise} the rotation frequency,
for every input model from our grids, by seeking its root.
Once the optimal rotation frequency $f_{\rm rot}^{\rm(opt)}$ is located, then
\begin{equation}\label{e-frot-opt}
\mathcal{M} = \mathcal{N}, \quad {\rm for} \quad f_{\rm rot}=f_{\rm rot}^{\rm (opt)}.
\end{equation}
A brief description of the algorithm that locates the root of $d\mathcal{N}$ is given in the 
Appendix\,\ref{ap-optim}.
Although we started from non-rotating models, we can optimise the rotation frequency, assuming rigid
rotation.
Since our proposed algorithm has minimum underlying assumptions, we propose that it can be applied to 
any slow to moderately rotating star, including the {\it Kepler} sample of $\gamma$\,Dor stars of 
\cite{vanreeth-2015-01,vanreeth-2015-02}, provided that we limit the study to the prograde and/or zonal modes. 
%This algorithm should not be applied to retrograde modes, since the TAR is invalid for those \citep{ballot-2010-01}.

% Figure %
\begin{figure}[t!]
\includegraphics[width=\columnwidth]{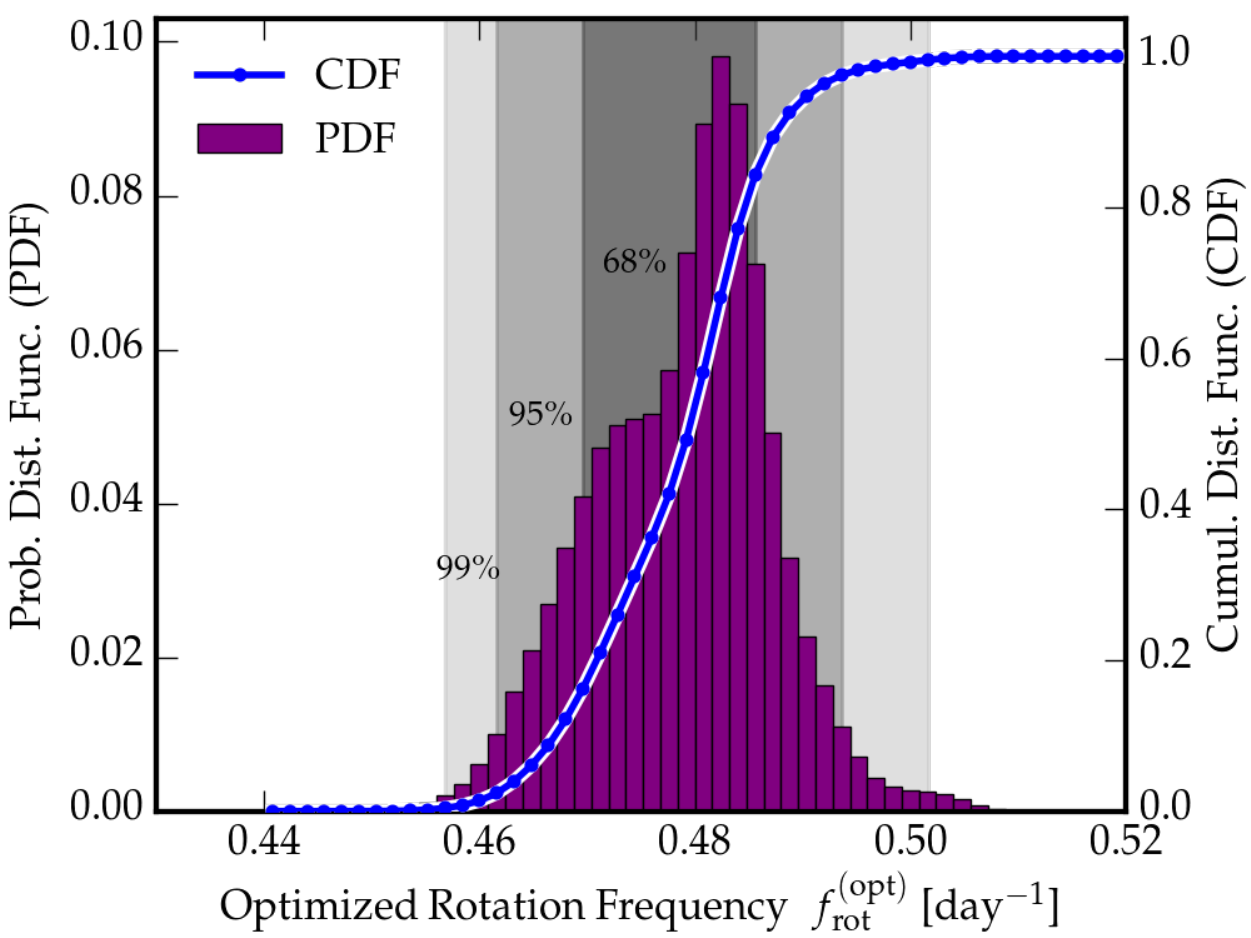} 
\caption{Histogram of the rotation frequency in grid A after rotation optimisation 
$f_{\rm rot}^{\rm(opt)}$ (see Eq.\,\ref{e-frot-opt}).
The blue curve is the cumulative distribution function (CDF) of the rotation frequency. 
The three shaded zones correspond to the rotation ranges where 99\%, 95\% and 68\% of $f_{\rm rot}^{\rm(opt)}$ 
lie, respectively.
\label{f-hist}}
\end{figure}

Fig.\,\ref{f-hist} shows the histogram of the optimised rotation frequency for all models in grid A, 
which can be interpreted as the probability distribution function (PDF) of $f_{\rm rot}$ by requiring 
all models to fulfil Eq.\,(\ref{e-frot-opt}).
The PDF exhibits a skewed distribution, strongly deviating from a Gaussian distribution.
The striking result of our optimisation scheme -- reflected in the PDF -- is that the 
$f_{\rm rot}^{\rm(opt)}$ for 99\% of the models lies between 0.4568 to 0.5016\,day$^{-1}$, which is
already a tight constraint on the possible rotation frequency of the target.
Similarly 68\% of rotation frequencies lie between 0.4696 to 0.4856\,day$^{-1}$.
The mean optimised rotation frequency $\langle f_{\rm rot} \rangle$ is simply a weighted average of optimised 
rotation frequencies $f_{{\rm rot},k}^{\rm (opt)}$ within each histogram bin $k$, with the PDF
within the same bin used as the weight $w_k$.
In other words, 
\begin{equation}\label{e-mean-frot}
  \langle f_{\rm rot} \rangle = \frac{ \sum_{k} w_k\, f_{{\rm rot},k}^{\rm(opt)} }{\sum_k w_k}. 
\end{equation}
Together with the 1$\sigma$ PDF range, the mean optimised rotation frequency of KIC\,7760680 is
$\langle f_{\rm rot} \rangle = 0.4790_{-0.0094}^{+0.0066}$\,day$^{-1}$.

At the end of this step, we append $f_{\rm rot}^{\rm (opt)}$ to every model in our grids, ensuring that
they reproduce the slope of the observed period spacing, and fulfil Eq.\,(\ref{e-dN}). 
Now, we can proceed to compute the $\chi^2$ goodness-of-fit scores (next section) to rank all our 
input models accordingly.

%%%%%%%%%%%%%%%%%%%%%%%%%%%%%%%%%%%%%%%%%%%%%%%%%%%%%%%%%%%%%%%%%%%%%%%%%%%%%%%%%%%%%%%%%%%%%%%%%%%%

\section{Model Frequencies and Ranking}\label{s-chisq}

Each equilibrium structure model from our grids is fed into the adiabatic linear nonradial pulsation
code GYRE, using TAR.
We compute dipole prograde frequencies within a broad trial range that ensures covering the observed 
period range.
In GYRE, the frequencies are internally computed in the co-rotating frame $f^{\rm(co)}_i$, 
for a trial rotation frequency $\omega_{\rm rot}=2\pi\,f_{\rm rot}$, but are stored in the inertial 
frame $f_i^{\rm(in)}$, considering the Doppler shift $f^{\rm(in)}_i=f^{\rm(co)}_i+m\,f_{\rm rot}$. 
This facilitates consistent comparison between the frequecies, periods and/or period spacings from 
observation and models.

Because we optimise the rotation frequency (Sect.\,\ref{s-optim-f-rot}), we can proceed to a
mode-by-mode comparison between observations and models.
To rank all models based on their quality of fitting the observed frequencies 
\citep[Table\,1][]{papics-2015-01}, we define a frequentist reduced-$\chi^2$ score, denoted by
$\chi^2_{\rm red}$
\begin{equation}\label{e-chisq}
%\chi^2_{\rm red} = \frac{1}{\mathcal{N}^{\rm(obs)}-1-n}\sum_{i=1}^{\mathcal{N}^{\rm(obs)}-1}
%    \left(\frac{ \Delta P_i^{\rm (obs)} - \Delta P_i^{\rm (mod)} }{\sigma(\Delta P)_i}\right)^2.
\chi^2_{\rm red} = \frac{1}{\mathcal{N}-n}\sum_{i=1}^{\mathcal{N}}
    \left(\frac{ f_i^{\rm (obs)} - f_i^{\rm (mod)} }{\sigma_i}\right)^2.
\end{equation}
where $\mathcal{N}=36$ is the number of observed modes, $n=5$ is the number of free 
parameters in each grid for the fixed input physics, and the $\sigma_i$ are the 1$\sigma$ 
uncertainties of observed frequencies.
Consequently, we sort and tabulate all input models based on their associated $\chi^2_{\rm red}$.
Based on this, we pick the two best model(s), one from grid A, and the other from grid B,
with minimum $\chi^2_{\rm red}$ scores. 
The properties of these models are discussed in Sect.\,\ref{s-results}.

%%%%%%%%%%%%%%%%%%%%%%%%%%%%%%%%%%%%%%%%%%%%%%%%%%%%%%%%%%%%%%%%%%%%%%%%%%%%%%%%%%%%%%%%%%%%%%%%%%%%

\section{Results}\label{s-results}

%%%%%%%%%%%%%%%%%%%%%%%%%%%%%%%%%%%%%%%%%%%%%%%%%%%%%%%%%%%%%%%%%%%%%%%%%%%%%%%%%%%%%%%%%%%%%%%%%%%%

\subsection{Best Asteroseismic Model Candidates}\label{ss-best}

\begin{figure*}[th!]
\includegraphics[width=\textwidth]{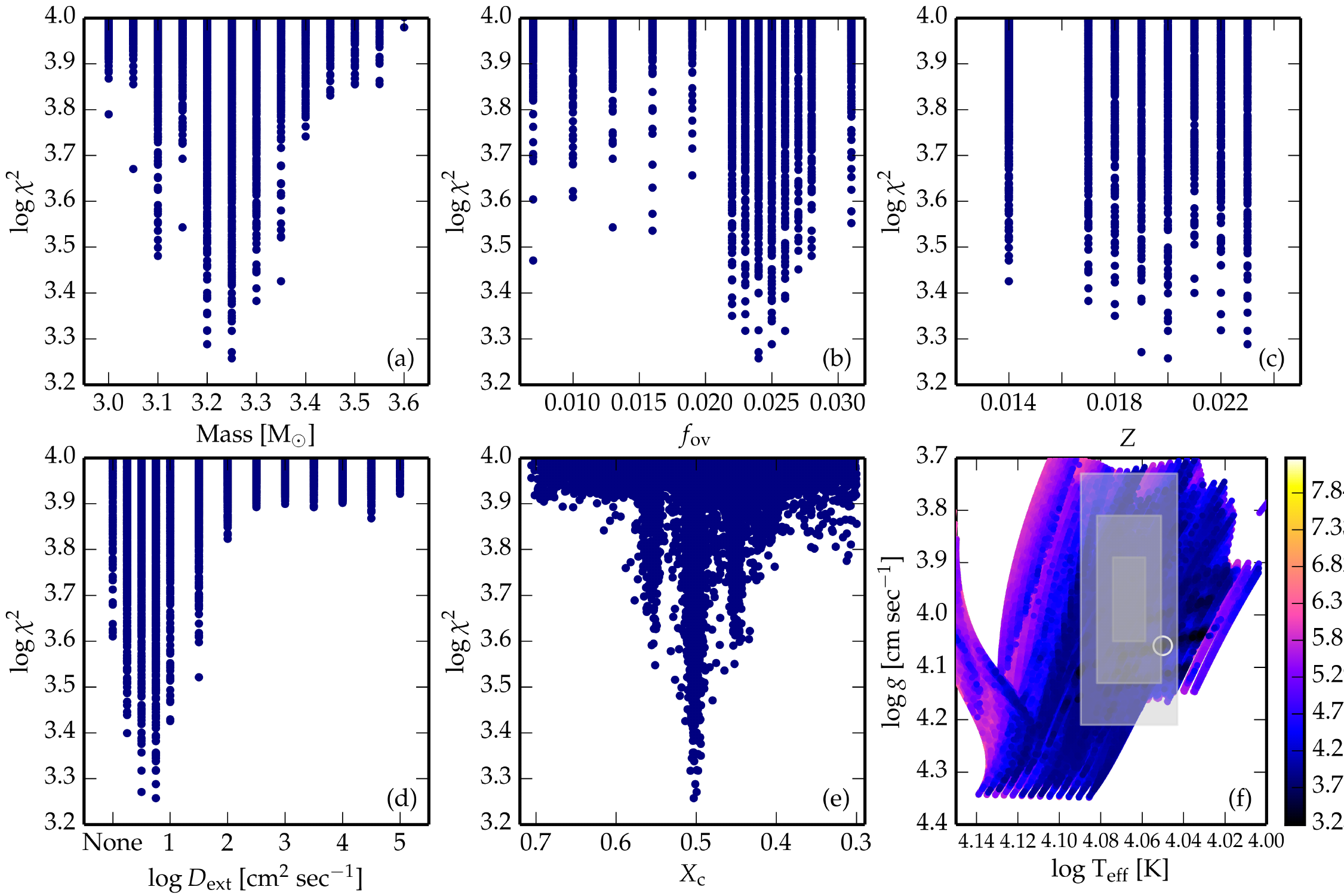} 
\caption{The distribution of the $\log\chi^2_{\rm red}$ (Eq.\,\ref{e-chisq}) for the free 
parameters of grid A.
A similar distribution for grid B is presented in Fig.\,\ref{f-chisq-aov}.
Panels (a) to (e) show the local minima of $\log\chi^2_{\rm red}$ versus initial mass M$_{\rm ini}$, 
exponential overshoot free parameter $f_{\rm ov}$, metallicity Z$_{\rm ini}$, extra diffusive mixing 
$\log D_{\rm ext}$, and center hydrogen mass fraction X$_{\rm c}$, respectively.
For clarity, the ordinate is restricted to models with $\log\chi^2_{\rm red}\leq4$.
Panel (f) shows the position of the input models on the Kiel diagram. 
The 1$\sigma$, 2$\sigma$ and 3$\sigma$ uncertainty boxes for the position of KIC\,7760680 from spectroscopy are
highlighted as grey boxes.
The position of the best model is flagged with a white open circle.
The $\log\chi^2_{\rm red}$ is colour coded.
\label{f-chisq-scatter}}
\end{figure*}

Panels (a) to (e) in Fig.\,\ref{f-chisq-scatter} show the distribution of the logarithm of 
$\chi^2_{\rm red}$ versus the free parameters of grid A.
The ordinate is limited to the lowest $\chi^2_{\rm red}$ values, despite 
$\chi^2_{\rm red}\in[1808, \,2.11\times10^8]$.
The $\log \chi^2_{\rm red}$ shows significant minima for most of the grid parameters, allowing to 
tightly constrain them.
Our best asteroseismic model corresponds to the one that has the minimum $\chi^2_{\rm red}$, 
within each grid.
In what follows, we call the best model from grid A as mA, and that of grid B as mB.
The internal structure of model mA is freely available for download, as explained in 
Appendix\,\ref{ap-files}.
Figs.\,\ref{f-Ov-Dmix}b and \ref{f-Ov-Dmix}c depict the internal structures of these two best models.
Table\,\ref{t-chisq} gives an overview of the physical properties of mA and mB that we 
elaborate below.

The large $\chi^2_{\rm red}$ values in Fig.\,\ref{f-chisq-scatter} is a common situation when modelling
heat-driven modes with such high precision frequencies.
In spite of that, the most plausible parameters of KIC\,7760680 are the following.
The initial mass is roughly M$_{\rm ini}=$3.25\,M$_\odot$, which is identical to that of 
Star\,\textsc{I}.
Thus, KIC\,7760680 is a moderately rotating analog of Star\,\textsc{I}.
There is a clear indication that Z$_{\rm ini}\approx0.020$, which agrees with the spectroscopic estimate 
that [M/H]=$0.14\pm0.09$.
Thus, KIC\,7760680 is a metal-rich dwarf.
From Fig.\,\ref{f-chisq-scatter}e, the age is well-constrained to X$_{\rm c}=0.50$, implying that 
KIC\,7760680 is still in its early main sequence evolution.
The best value for the exponential overshooting parameter is $f_{\rm ov}=0.024$, although 
$f_{\rm ov}\in[0.022, \,0.026]$ result in comparatively good fits to the observed period spacing. 
Thus, the overshooting is stronger in this target, compared to that of the slower rotator 
Star\,\textsc{I}.

The extra diffusive mixing exhibits a distinct minimum around $\log D_{\rm ext}\approx0.75$. 
Theoretical predictions for the vertical (radial) component of the shear-induced mixing $D_v$
in differentially rotating massive stars is roughly three to even ten orders of magnitude stronger 
than what we constrained here;
for several examples refer to Fig.\,7 in \cite{talon-1997-01}, Fig.\,6 in \cite{meynet-2000-01},
Fig.\,3 in \cite{mathis-2004-01}, and Figs.\,15 and 16 in \cite{decressin-2009-01}.
Based on Eq.\,(7) in \cite{mathis-2004-01}, $D_v$ depends explicitly on the square of the 
angular differential rotation frequency $D_v\propto (r\,d\Omega/dr)^2$.
The immediate -- and perhaps most plausible -- explanation of the low $D_{\rm ext}$ value is that 
KIC\,7760680 is nearly a rigid-body rotator.
The range of viable $D_{\rm ext}$ is so negligibly small that neglecting additional mixing in the 
radiative envelope is justifiable for this star.
The extent of the three mixing regions, in addition to the profile of the hydrogen abundance and 
the Brunt-V\"ais\"al\"a frequency for the best model are shown in Figs.\,\ref{f-Ov-Dmix}a and 
\ref{f-Ov-Dmix}b.

% Figure %
% Figure %

Fig.\,\ref{f-chisq-scatter}f shows the position of all evolutionary tracks on the Kiel diagram.
The color coding is proportional to $\log \chi^2_{\rm red}$. 
The 1$\sigma$, 2$\sigma$, and 3$\sigma$ uncertainty boxes are highlighted in grey colour.
The position of the best model is flagged with a white circle, which confirms that the asteroseismic
parameters of the best model agree with their spectroscopic counterparts;
recall that the same agreement was already achieved between $v\sin i$ and the optimised rotation frequency.
%The branch of evolutionary tracks that evolve to higher effective temperatures correspond to those models
%with high overshooting and extra mixing in their envelope, so that they evolve chemically homogeneous.

% Table
\begin{table}[t]
\caption{
Deduced physical properties of two best asteroseismic models from grid A and B (in Table\,\ref{t-grids}).
%The two grids grids have different overshoot prescriptions.
Uncertainties in grid parameters are set by the minimum stepsize in Table\,\ref{t-grids},
and given in parentheses, which is a lower limit of the true uncertainty.
Refer to Figs.\ref{f-Ov-Dmix}b and \ref{f-Ov-Dmix}c for the extent of each mixing 
region in both models.
}
\label{t-chisq}
\begin{center}
\begin{tabular}{lll}
  \hline
%  Parameter & Fine & Step-Function \\
%            & grid & grid \\
  Parameter & & \\
  \hline
  Model Name& mA & mB \\
  Grid      & A & B \\
  $\chi^2_{\rm red}$ & 1808 & 3647 \\
  M$_{\rm ini}$ [M$_\odot$] & 3.25\,(5) & 3.00\,(5) \\
  Z$_{\rm ini}$ & 0.020\,(1)  & 0.023\,(1) \\
  Overshoot & $f_{\rm ov}$=0.024\,(1) & $\alpha_{\rm ov}$=0.32\,(1) \\
  $\log D_{\rm ext}$ [cm$^2$\,sec$^{-1}$] & 0.75\,(25) & 0.50\,(25) \\
  X$_{\rm c}$  & 0.503\,(1)  & 0.496\,(1) \\
  $f_{\rm rot}^{\rm(opt)}$ [day$^{-1}$] & 0.4805 & 0.4744 \\
  $f_{\rm rot}^{\rm(opt)}/f_{\rm crit}$ [\%] & 26.4 & 26.6 \\
  \hline
  M$_\star$ [M$_\odot$]    & 3.2499 & 3.0000 \\  
  R$_\star$ [R$_\odot$]    & 2.7895 & 2.7501 \\  
  L$_\star$ [L$_\odot$]    & 110.8  & 75.0 \\  
  Age [10$^6$ yr]          & 202    & 278 \\
  \hline
  $m_{\rm cc}$ [M$_\odot$] & 0.6215 & 0.5437 \\
  $r_{\rm cc}$ [R$_\odot$] & 0.3356 & 0.3109 \\
  $m_{\rm ov}$ [M$_\odot$] & 0.2642 & 0.2239 \\
  $r_{\rm ov}$ [R$_\odot$] & 0.0558 & 0.0495 \\
  \hline
\end{tabular}
\end{center}
\end{table}
% Table

That the resulting $\chi^2_{\rm red}$ are larger than one thousand (even for the best models) 
stems from two facts.
First, the relative uncertainties in the mode frequencies $\sigma_i/f_i$ 
\citep[see e.g. Fig.\,\ref{f-obs-dP}b or Table\,1 in][]{papics-2015-01} are roughly $\sim10^{-4}$
to $10^{-6}$.
Second, our current understanding of stellar structure and evolution is based on 1D models, imposing 
simplifying assumptions (e.g. stellar opacity, stellar composition and mixture, clumpiness in mass loss, 
treatment of rotation), ignoring some physical processes (e.g. atomic diffusion, radiative levitation,
magnetic field), in addition to other physical processes that are not understood well (e.g. 
the role of internal gravity waves, the angular momentum transport, interaction of various mixing
processes).
Therefore, it is not surprising that our 1D equilibrium models succeed to explain the overall 
asteroseismic observables globally, but not in detail.

\begin{figure*}[t]
\includegraphics[width=\textwidth]{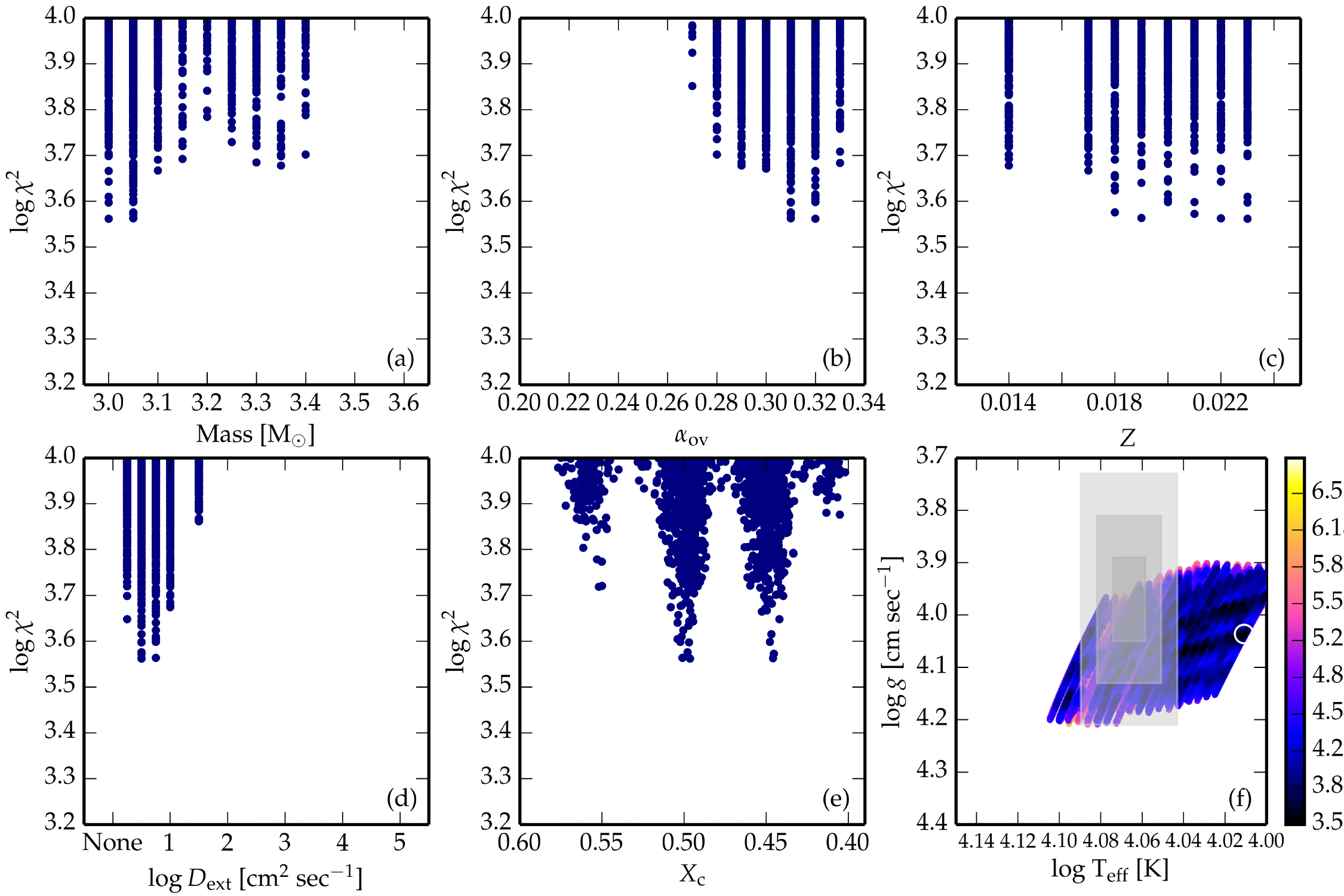} 
\caption{Similar to Fig.\,\ref{f-chisq-scatter}, but for the grid B with step-function overshooting.
The ordinate of the two figures are identical.}
\label{f-chisq-aov}
\end{figure*}

The distribution of $\log\chi^2_{\rm red}$ for grid B (with step-function overshoot) is presented in 
Fig.\,\ref{f-chisq-aov}, and the grid parameters are given in Table\,\ref{t-grids}.
The $\log\chi^2_{\rm red}$ values lie in the range [3.562, 6.825].
The preferred value for step-function overshoot is $\alpha_{\rm ov}=0.31$ to 0.32.
Thus, as with grid A, grid B indicates that sizeable overshoot mixing is required to match the 
observations.
Furthermore, the extra diffusive mixing is also constrained to $\log D_{\rm ext}=0.50$ to 0.75,
in excellent agreement with $\log D_{\rm ext}=0.75$ found in grid A.
In contrast with grid A (Fig.\,\ref{f-chisq-scatter}), constraining the initial mass, metallicity and age from 
grid B is less conclusive, and a broader range of values provide equally good overall frequency fits.
Fig.\,\ref{f-Ov-Dmix}c shows the mixing property of the best model in grid B.

% Figure
% Figure

\begin{figure*}[t]
\includegraphics[width=\textwidth]{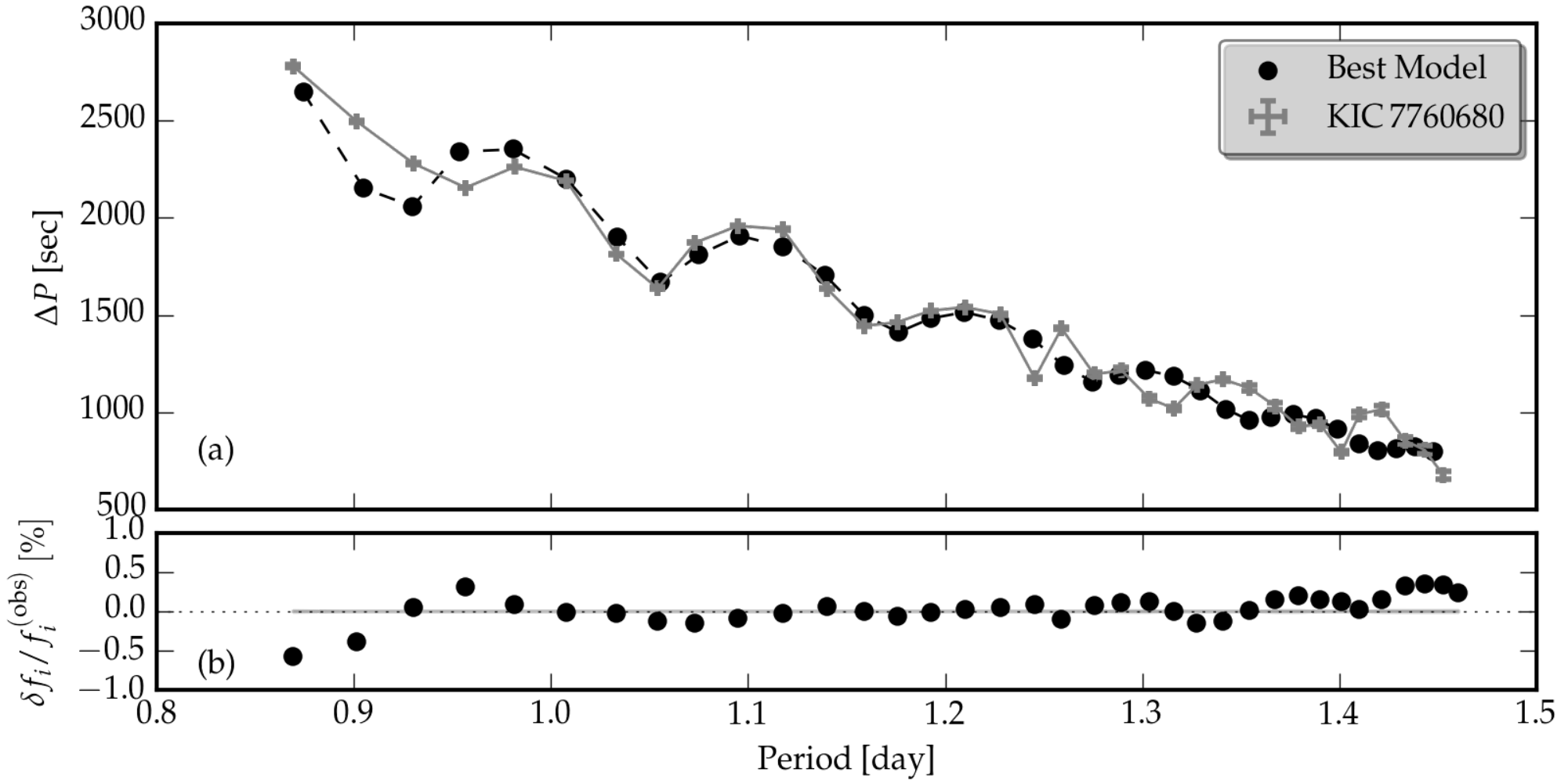} 
\caption{(a) Comparing period spacing from observation (grey) versus best-fit model (black) from 
grid A.
The parameters of this model are given in Table\,\ref{t-chisq}.
For a comarison with Star\,\textsc{I}, refer to Fig.\,4 in \cite{moravveji-2015-01}.
The two patterns reasonably match up to $\sim1.23$\,day$^{-1}$, beyond which the two start to deviate.
(b) Percentage of relative frequency difference between the observation and the model,
$\delta f_i/f_i^{\rm (obs)}$.
The narrow grey area in the middle is the observed 1$\sigma$ uncertainty shown in Fig.\,\ref{f-obs-dP}b.
Note the presence of a cyclic pattern in $\delta f_i/f_i^{\rm (obs)}$.
\label{f-best-dP}}
\end{figure*}

Fig.\,\ref{f-best-dP}a shows the resulting period spacing of ``the best model" (black
filled circles) from mA, compared to the observations (grey symbols, from Fig.\,\ref{f-obs-dP}).
The overall fit to the slope of the period spacing is excellent, thanks to the optimised rotation 
frequency $f_{\rm rot}^{\rm (opt)}$. 
More importantly, the local dips in the observed period spacing are reasonably reproduced,
mainly for the shorter-period modes.
Towards the longer-period end (with increasing radial order), the quality of the fit noticeably degrades.
Fig.\,\ref{f-best-dP}b shows the relative frequency difference between models and observations
$\delta f_i/f_i^{\rm (obs)}$, where $\delta f_i=f_i^{\rm(obs)} - f_i^{\rm(mod)}$.
All values are below 1\%, so the agreement between observed and theoretical frequencies is at this level.
The presence of the strong cyclic pattern in $\delta f_i$, among other things, hints at the possible 
presence of a glitch in Brunt-V\"ais\"al\"a frequency which is not accounted for in our current 
treatment of the overshooting mixing.
The reason for this, as already elaborated above is the inadequacy of our knowledge about the 
convective boundary mixing by overshooting, and the missing physics in our 1D models that render the 
thermal and chemical stratification above the fully mixed core.

% Figure %
% Figure %

The inferred mean rotation frequency for mA is 26.4\% of the Roche critical 
rotation frequency $f_{\rm crit}^{\rm (Rch)}$, which is defined as 
\begin{equation}\label{e-f-crit}
f_{\rm crit}^{\rm (Rch)} = \frac{1}{2\pi} \left(\frac{8}{27} \frac{G\,M_\star}{R_\star^3}\right)^{1/2},
\end{equation}
where $G$, M$_\star$ and R$_\star$ are respectively the gravitational constant, stellar mass and radius 
-- from the best non-rotating 1D model.
KIC\,7760680 is thus a moderately rotating SPB star.
The corresponding estimate of the mean equatorial rotation velocity 
$v_{\rm eq}=2\pi R_{\star}<f_{\rm rot}>\approx63$ km\,sec$^{-1}$, which agrees remarkably with the 
projected rotation velocity $v\,\sin i=62\pm5$ km\,sec$^{-1}$.
This means that the inclination angle is $i\approx80^\circ$, and KIC\,7760680 is observed nearly equator on,
totally compatible with the detected sectoral ($m\neq0$) modes.

Previously, we demonstrated that the exponentially decaying overshoot prescription provided a superior
frequency fit to the observed modes of Star\,\textsc{I} \cite[see Table\,3 in][]{moravveji-2015-01}.
The reason was that the corresponding $\chi^2_{\rm red}$ for the best step-function overshoot model 
was roughly 2.2 times worse (larger) than that from exponentially-decaying overshoot grid.
We repeat the same exercise here, thanks to the extensive Step-Function grid (B) we computed 
(Table\,\ref{t-grids}).
The $\chi^2_{\rm red}$ for the step-function overshoot model from Table\,\ref{t-chisq} is  
$\sim$ twice the one of the exponential overshoot.
This is the second case for which the exponential overshoot 
prescription is favoured over the classical step-function overshoot.

Table\,\ref{t-chisq} summaries the physical properties of the two best models from grids A and B.
Despite different $\chi^2_{\rm red}$ scores, their associated parameters posses interesting 
similarities, even when adopting two different overshooting prescriptions.
The initial mass and metallicity of both models are found reasonably close;
the lower-mass model has higher metallicity, and vice versa.
This is the well-established M$_{\rm ini}$-Z$_{\rm ini}$ correlation, also shown in 
\cite{ausseloos-2004-01} for \objectname{$\nu$\,Eri}, in \cite{briquet-2007-01} for 
\objectname{$\theta$\,Oph}, and in \cite{moravveji-2015-01} for Star\,\textsc{I}.
The core hydrogen content X$_{\rm c}$, and rotation frequency are found to be consistently close in 
both models.
Most importantly, very low extra diffusive mixing $D_{\rm ext}$ is required in both models.
Therefore, the global properties of the star, inferred from our seismic modelling, weekly depend 
on the adopted overshooting model.
Regardless of the overshoot prescription option, both models point at
a sizeable amount of overshoot, in the presence of rotation.
This strengthens our earlier conclusion that moderate rotation requires larger core overshooting, 
than slow rotation does.

In Table\,\ref{t-chisq}, we give the mass contained in the fully mixed convective core $m_{\rm cc}$,
the mass contained in the overshooting region $m_{\rm ov}$, and their radial extents $r_{\rm cc}$ and 
$r_{\rm ov}$, respectively, from the profiles of the equilibrium structure.
The relative extent of the overshooting regions in both models (Table\,\ref{t-chisq}) are almost 
identical.
The relative overshooting mass with respect to the star mass $m_{\rm ov}/M_\star$ in mA (or mB) 
is 8.1\% (or 7.5\%).
In terms of radial extent, $r_{\rm ov}/R_\star$ for the mA (or mB) is 
2.0\% (or 1.8\%).
Although, the overshooting region is a narrow part of the star, we show in Sect.\,\ref{ss-trap} that
many high-order g-modes are perfectly trapped inside this layer.
Similarly, we can compare the relative mass and radial extent of the overshooting zone with respect
to that of the convective core to shed light on the distance over which the convective eddies travel
before losing their identity and falling back.
For mA (or mB), $m_{\rm ov}/m_{\rm cc}$ is 42.5\% (or 41.2\%), 
and similarly $r_{\rm ov}/r_{\rm cc}$ is 16.6\% (or 15.9\%).
Therefore, the fully mixed cores in late B-type stars require $\sim16\%$ increase in mass beyond their
canonical boundary from the MLT.
We argue that future more advanced non-local time-dependent theories of convection 
\citep{xiong-1979-01,xiong-1989-02,canuto-2011-01,canuto-2011-05,zhang-2012-03,zhang-2013-01,
zhang-2016-01,arnett-2015-01} should closely reproduce our seismic findings.
Three dimensional simulations by \cite{browning-2004-01} for rotating A-type stars already 
predicted that the convective cores require at least $d_{\rm ov}\gtrsim0.20\,H_p$ overshooting 
extension beyond the canonical boundary, depending on the stiffness (buoyancy jump) of the 
stratification.
The need for such core extension in our results agrees with the predictions of 
\citeauthor{browning-2004-01}

The relative frequency deviations $\delta f_i/f_i^{\rm(obs)}$ for model mA is 
below one percent.
This is clearly shown in Fig.\,\ref{f-best-dP}b.
For Star\,\textsc{I}, this was below 0.3 percent, a factor three better than KIC\,7760680.
Therefore, our seismic models serve as perfect starting point for frequency inversion of the 
gravity modes, in order to improve the assumed thermal and chemical stratification of the 
overshooting, beyond the current available models.
Although the structure inversion theory for solar-type p-mode pulsators is well established
\citep[see][and references therein]{basu-2014-01,buldgen-2015-02}, no such theory is yet developed
for heat-driven high-order g-modes in massive star.

%%%%%%%%%%%%%%%%%%%%%%%%%%%%%%%%%%%%%%%%%%%%%%%%%%%%%%%%%%%%%%%%%%%%%%%%%%%%%%%%%%%%%%%%%%%%%%%%%%%%

\subsection{Nonadiabatic Mode Stability}\label{ss-non-ad}

\begin{figure}[t]
\includegraphics[width=\columnwidth]{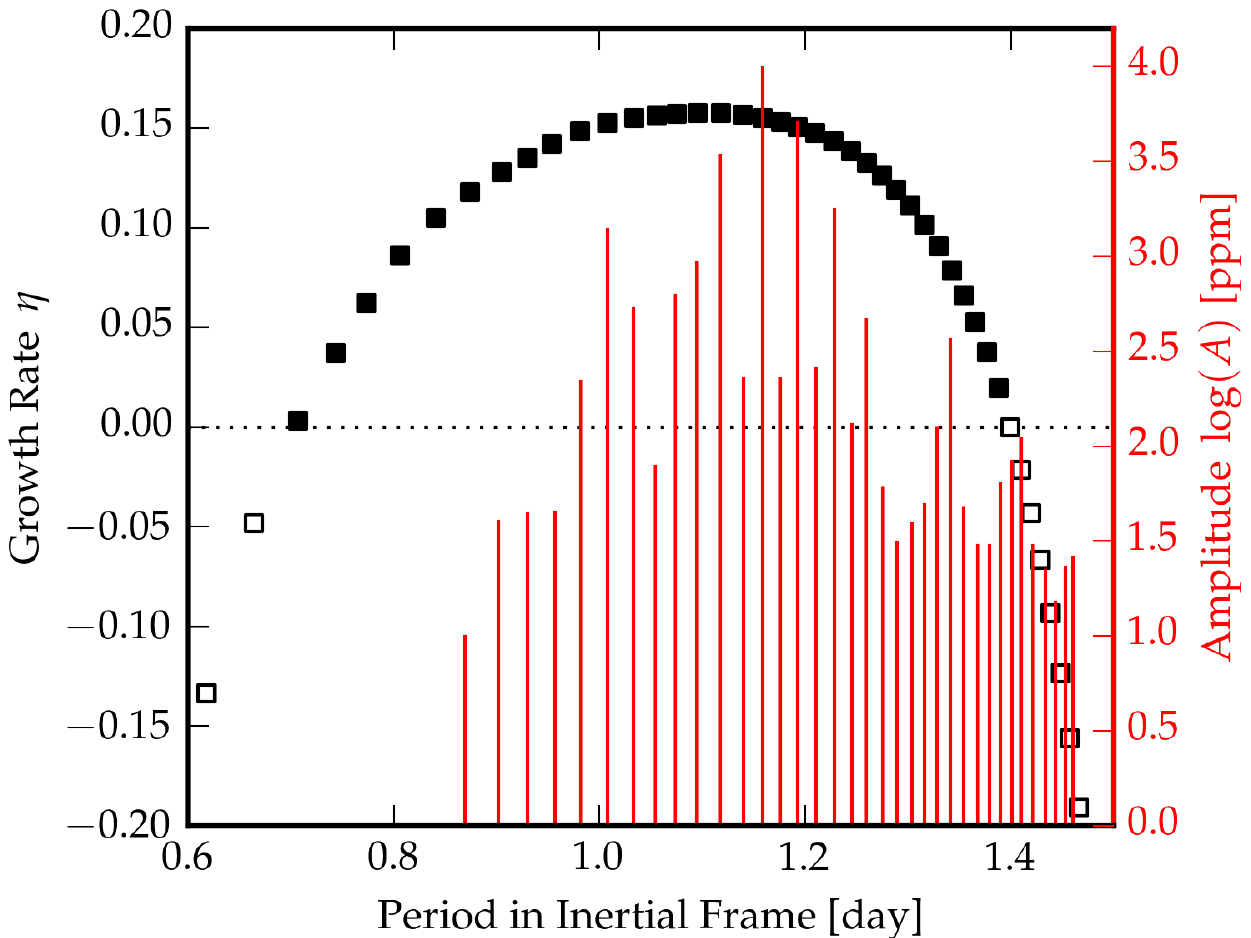} 
\caption{Normalised growth rates $\eta$ for the best model.
The unstable (stable) modes are presented with black filled (empty) squares.
The logarithm of the observed mode amplitude $A$ of the dipole series is shown with vertical red 
lines.}
\label{f-growth}
\end{figure}

As an a posteriori test, we consider mode stability properties by computing the non-adiabatic 
dipole prograde frequencies \citep{townsend-2005-01,townsend-2005-02}.
For the best model, the radial order of the modes that match the observation lie in the range
$-53\leq n_{\rm pg} \leq -18$.
Fig.\,\ref{f-growth} shows the normalised growth rate $\eta=W/\int_{0}^{R_\star}|dW|$ for the best 
model, as first introduced by \cite{stellingwerf-1978-02}. 
Here, $W$ is the total work, from integrating the work integrand $dW$ over the whole star 
$W=\int_{0}^{R_\star}dW$.
Unstable (or stable) modes correspond to positive (or negative) $\eta$ values, and are shown with 
filled (or empty) squares.  
From $\mathcal{N}=36$ modes, a total of 34 modes are predicted to be unstable, which 
is in excellent agreement with observations, thanks to employing OP tables with enhanced Fe and 
Ni monochromatic opacities.
Ignoring the important role of Fe and Ni drastically underestimates the predicted excited modes
\citep{dziembowski-2008-01,salmon-2012-01}.
Therefore, in addition to solving the $\beta$\,Cep and hybrid pulsating massive stars excitation 
problem presented in \cite{moravveji-2016-01}, the success in explaining the excitation of the 
majority of observed modes by incorporating Fe and Ni monochromatic opacity enhancement is another 
manifestation that the default (OP and OPAL) opacity tables underestimate the Rosseland mean opacity 
in stellar interiors.
Consequently, the stellar interior seems more opaque than believed, and next generation of stellar
models should adopt updated tables of \cite{moravveji-2016-01} and/or \cite{mondet-2015-01}.

In Fig.\,\ref{f-growth}, the period instability domain does not perfectly agree with observations, 
and seems shifted towards lower-order modes:
five short-period g-modes are predicted unstable, but are not observed, in addition to 
seven long-period g-modes that are observed, but predicted to be stable.
We previously found similar issue for Star\,\textsc{I};
see Fig.\,9 in \cite{moravveji-2015-01}.

% Figure %
% Figure %

We speculate that the slight mismatch in Fig.\,\ref{f-growth} between observed and predicted excited modes 
can be possibly explained by any of, or the combination of the following three missing physical 
inputs in the current 1D evolutionary models.
(1) Firstly, the gravitational settling and radiative levitation are ignored in our 1D models 
    (due to their $\sim$100 times longer computation overhead). 
    The 3D and 1D simulations of \cite{zemskova-2014-01} for a 1.5\,M$_\odot$ star showed that 
    Fe can gradually accumulate in the iron-bump, modifying the local metallicity, without noticeable 
    change of the surface metallicity.
    With KIC\,7760680 having more than twice the mass contained in the simulations of \citeauthor{zemskova-2014-01}, 
    the radiative levitation can dominate even further, and contribute very efficiently to iron and 
    nickel accumulation around the iron bump.
    This important feature is still missing from our MESA models.
(2) Secondly, the Fe and Ni are the major contributors to the iron opacity bump, whose abundance in 
    KIC\,7760680 are assumed to be solar. 
    This may not necessarily be true.
    Thus, a slight increase in the Fe and Ni initial abundance (at the cost of slight reduction in 
    initial hydrogen and/or helium) can potentially resolve this problem.
    This is beyond our current scope, because it calls for re-computing (even a part of) our 
    asteroseismic grid for unknown initial chemical mixtures $X_i$, for $i=$H, He, $\cdots$, Fe, Ni.
(3) Thirdly, the iron opacity peak occurs around $\log T\approx5.2$ to 5.3\,dex.
    We speculate a slight inward shift of the opacity peak towards the hotter interior can help overcoming the 
    radiative damping, and alleviate the lack of sufficient excited modes.

%%%%%%%%%%%%%%%%%%%%%%%%%%%%%%%%%%%%%%%%%%%%%%%%%%%%%%%%%%%%%%%%%%%%%%%%%%%%%%%%%%%%%%%%%%%%%%%%%%%%

\subsection{Mode Trapping in The Deep Stellar Interior}\label{ss-trap}

It is instructive to consider the modal behaviour (of our best model) to demonstrate the probing 
power of high-order g-modes in the deep stellar interior.
The rotational kernels $\mathcal{K}_{n,\ell}$, and mode inertia $\mathcal{I}_{n,\ell}$ --
which are constructed from the radial and horizontal components of eigendisplacements and defined
in \cite{aerts-2010-book} -- are two useful quantities to exploit.
The kernels of high-order g-modes become progressively oscillatory by the increase in 
mode radial order (and mode period), and attain larger amplitude towards the core, compared to the
surface.
This makes high-order g-modes in rotating SPB stars ideal probes of the near-core environment, 
provided that the local wavelength of the mode is roughly equal to or smaller than the length 
scale of the change of structure in the background model \citep{cunha-2015-01,belyaev-2015-01}.
In such cases, the model g-modes are able to {\it resolve} the structure of their background 
medium (which they propagate in), and their frequencies reveal the shortcomings in treating the 
near-core thermal and chemical stratification, by deviating from observations.
We argue that a subset of g-modes in KIC\,7760680 are trapped inside the overshooting region, and reveal that
the current state of the modelling of chemical mixing and thermal stratification in that region 
are not accurate enough to explain the high precision data.

\begin{figure*}[th!]
\begin{minipage}{0.5\textwidth}
\includegraphics[width=\columnwidth]{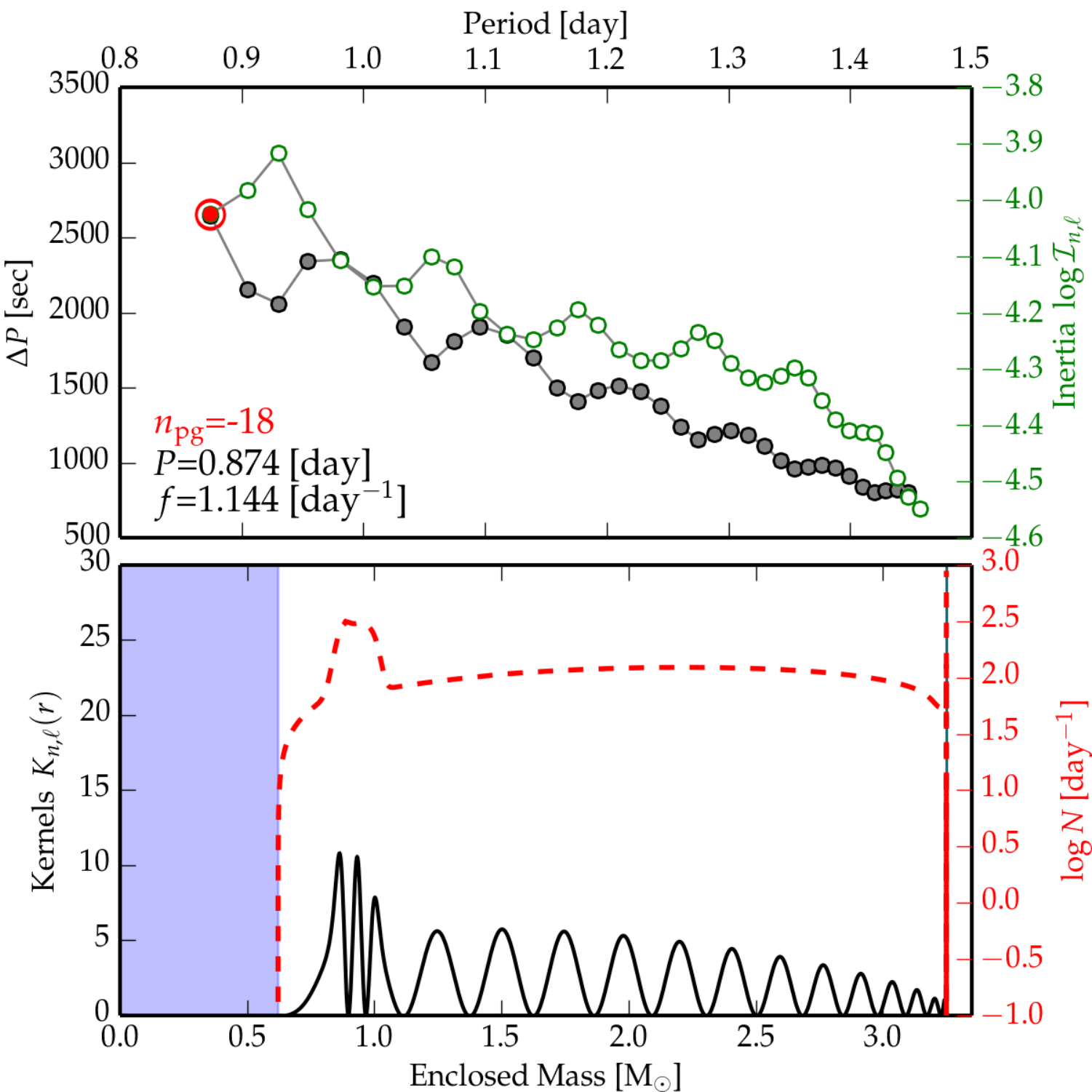}
\end{minipage}
\begin{minipage}{0.5\textwidth}
\includegraphics[width=\columnwidth]{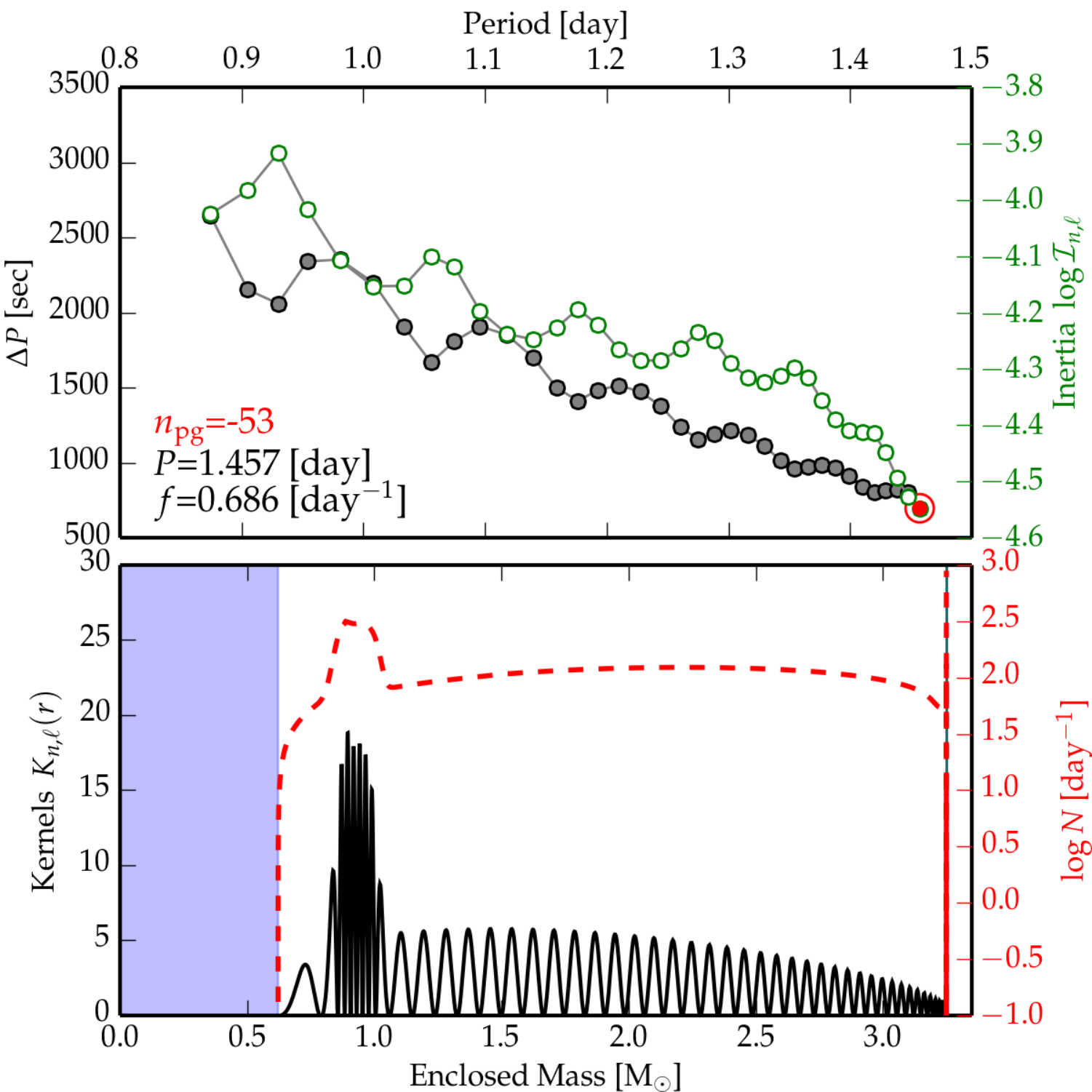} 
\end{minipage}
\caption{The mode Kernels $\mathcal{K}_{n,\ell}$ (bottom solid line) and the logarithm of mode inertia 
$\mathcal{I}_{n,\ell}$ (top empty dots) for the shortest-period mode with period $P_1=0.873$\,d$^{-1}$ 
(left), and for the longest-period mode with period $P_{36}=1.457$\,d$^{-1}$ (right). 
On the top panels, the period spacing is shown in grey filled dots.
The red empty circles mark the corresponding mode.
On the bottom panels, the blue area shows the convective core where g-modes are evanescent, and 
the red dashed line shows the Brunt-V\"ais\"al\"a frequency $\log N$.
The best model from Table\,\ref{t-chisq} is used as the input.
\label{f-Knl}}
\end{figure*}

The two panels in Fig.\,\ref{f-Knl} compare several seismic properties of the lowest-order g-mode
$n_{\rm pg}=-18$ (left), and the highest-order one $n_{\rm pg}=-53$ (right) in the best model.
They represent the two extreme mode behaviours in the observed series, while
those of the intermediate modes exhibit a smooth transition between the two shown here.
The top panels show the period spacing (filled dots) and mode inertia (empty dots) versus mode periods.
The bottom panels show the profile of normalised rotational kernels $K_{\rm n,\ell}$;
in this panel, the convective zone is highlighted in blue, and the Brunt-V\"ais\"al\"a frequency
is shown with a dashed red line.

% Figure %
% Figure %

The kernels exhibit two significant features on top of the convective core: 
(a) Both modes exhibit partial trapping in the $\mu$-gradient region -- associated with the broad 
    bump in the Brunt-V\"ais\"al\"a frequency.
    The kernels of the modes associated with the dips in the period spacing attain highest
    relative amplitude in the $\mu$-gradient region, and become fully trapped.    
(b) The kernels of the modes with radial order exceeding $\sim30$, i.e. $n_{\rm pg}\lesssim-30$, 
    exhibit an additional trapping in the overshooting region between the boundary of the convective 
    core, and the base of the bump in the Brunt-V\"ais\"al\"a frequency.
    Notice the final emergence of a fully-trapped mode in the bottom left panel in Fig.\ref{f-Knl}, 
    at the mass coordinate $m\approx0.75$\,M$_\odot$.
    The highest-order modes which exhibit this additional trapping are those which probe the 
    overshooting zone, and provide very precise diagnostic of the physical structure of this region.

Consequently, the entire series of identified g-modes of KIC\,7760680 allow exploiting the extent and 
physical conditions of chemically homogeneous (overshooting) and inhomogeneous ($\mu$-gradient) 
layers in massive stars.
The success (or failure) in matching the individual observed frequencies is a reward (or penalty) 
of the accuracy of our current understanding and implementation of the physics of stellar interior.
The fact that lower-order modes ($n_{\rm pg}\gtrsim-30$) better match the observation is a clear 
evidence that the structure of the $\mu$-gradient layer is well represented in our 1D evolutionary 
model.
However, the evident period spacing deviations of higher order modes ($n_{\rm pg}\lesssim-30$) 
from observations unambiguously indicates a lack of missing physics of the overshoot mixing.
This explains the cyclic deviations between frequencies of the best model from observations presented
in Fig.\,\ref{f-best-dP}b.
This can be attributed to the ad-hoc implementation of overshoot (e.g. Sect.\,\ref{s-mix}) in 
one-dimensional models, since our local, time-averaged description of convective mixing by MLT
does not consistently account for convection-induced mixing beyond the core boundaries
\citep[e.g.][and references therein]{browning-2004-01,arnett-2014-01}.

The non-local, time-dependent convective models \citep[see e.g. ][]
{xiong-1979-01,xiong-1989-02,canuto-2011-01,canuto-2011-05,zhang-2012-03,zhang-2013-01,zhang-2016-01,
arnett-2015-01, pasetto-2014-01, pasetto-2015-01}
may be able to provide better insight into the physical structure of the overshoot region, and allow
improving the fit to the observed period spacing of g-mode pulsators.
Therefore, the long series of dipole period spacing in KIC\,7760680 and Star\,\textsc{I} provide two ideal tests
for theories of convective and non-convective heat, chemical and angular momentum transport.

%%%%%%%%%%%%%%%%%%%%%%%%%%%%%%%%%%%%%%%%%%%%%%%%%%%%%%%%%%%%%%%%%%%%%%%%%%%%%%%%%%%%%%%%%%%%%%%%%%%%

\section{Summary, Discussion and Conclusion}\label{s-end}

Let us revisit and discuss the four questions raised in Sect.\,\ref{s-intro}.
In this paper, we carried out a thorough forward seismic modelling of KIC\,7760680, the richest SPB star 
discovered so far.
We computed two non-rotating MESA grids, and incorporated the effect of rotation on g-modes by 
employing the traditional approximation. 
The unknown equatorial rotation frequency of the target was varied and optimized by enforcing to 
match the exact number of dipole prograde modes within the observed range; 
this approach, automatically reproduces the negative slope in the observed period spacing.
All models in our grids were ranked by a $\chi^2_{\rm red}$ merit function that accounted for  
fitting of the frequencies.
We showed that KIC\,7760680 is a 3.25\,M$_\odot$ SPB star that rotates at $\sim26\%$ its Roche breakup 
frequency.
At this moderate rotation velocity, substantial overshooting is required to match the frequencies.
Therefore, we demonstrate that rotation increases convective overshooting from the core, compared to the 
non-rotating case.

Considering Star\,\textsc{I}, we managed to tightly constrain $f_{\rm ov}$ to $\sim0.017$, and 
favored the exponential prescription over the step-function.
It is noteworthy that the diffusive exponential prescription is also supported by the time-dependent 
convection model of \cite{zhang-2013-01}, \cite{zhang-2012-02}, and \cite{zhang-2016-01} (and the 
references therein), which contradicts with the predictions of \cite{zahn-1991-01} and 
\cite{viallet-2015-01} that core overshooting results into an adiabatic extension of the core.
Our seismically derived overshoot value for the best models are $f_{\rm ov}=0.024$ and $\alpha_{\rm ov}=0.32$.
They are in excellent agreement with the previous studies 
\cite[reviewed by][and the references therein]{aerts-2013-01}, in addition to those of 
\cite{stancliffe-2015-01} 
from fitting the global observables of nine binary systems, between 1.3 and 6.2\,M$_\odot$, 
and the seismic modelling of \cite{deheuvels-2016-01} for F stars at the onset of having convective cores.
Thus, a global picture on the overshoot mixing for a broad range of stellar masses is gradually emerging.

We allowed substantial mixing in the radiative regions as we chose to vary $D_{\rm ext}$ from zero 
up to $10^5$. 
We found at least an order of magnitude smaller values, compared to the theoretical predictions of
\cite{mathis-2004-01}.
In case shear-induced turbulent mixing were important in KIC\,7760680, then our seismic models -- 
through their $\chi^2_{\rm red}$ scores -- should have preferred higher values for $\log D_{\rm ext}$.
From Fig.\,\ref{f-chisq-scatter}d, this is obviously not the case.
Since the shear-induced mixing depends explicitly on the gradient of angular velocity, we infer that 
such gradient is small, if not zero, and the upper limit of the resulting effective diffusion transport 
coefficient is roughly $D_{\rm ext}\lesssim10$. 
The absence of differential rotational, and shear-induced mixing is a strong evidence that
KIC\,7760680 is most probably rotating rigidly.
This is quite acceptable in the light of finding small shear discovered in two {\it Kepler} F-type 
stars \citep{kurtz-2014-01,saio-2015-01}. 
In fact, heat-driven g-modes \citep{lee-1993-01} and stochastically excited internal gravity waves
\citep{rogers-2015-01} are predicted to efficiently redistribute angular momentum inside 
B-type stars, and induce near-rigid rotation.
KIC\,7760680 could be another manifestation of this.
In the meanwhile, we cannot exclude other possible mechanisms that can suppress mixing, and enforce
rigid body rotation.
In the light of recent advancements in asteroseismology, deep-rooted fossil magnetic fields 
turn out to be ubiquitous in intermediate main sequence stars \citep{stello-2016-01}, and their
red giant descendants \citep{fuller-2015-01}.
Asteroseismic modelling of the the magnetic $\beta$\,Cep pulsator V\,2052\,Ophiuchi by 
\cite{briquet-2012-01} revealed that an internal magnetic field can suppress core overshooting.
In addition, \cite{briquet-2016-01} showed that the weakly magnetic B2\,IV-V SPB star $\zeta$\,Cas 
rotates rigidly, with a magnetic field of strength 100--150\,G inhibiting mixing in its envelope.
The observed nitrogen enhancement was then attributed to the transport by internal gravity waves.
%Therefore, the presence of fossil magnetic field in intermediate stars \citep{stello-2016-01} can 
%have profound impacts on the internal structure of our models, among which is inhibition of mixing in the
%radiative interior \cite{mathis-2005-01,briquet-2016-01}.
Because KIC\,7760680 is an intermediate mass star, it can be a showcase of magnetic inhibition of chemical mixing
by rigid rotation.

We inferred (in Sect.\,\ref{ss-best}) that KIC\,7760680 is nearly a solid body rotator, because the extra 
diffusive mixing is limited to $\log D_{\rm ext}\approx0.75$.
This is not surprising.
The two $\gamma$\,Dor stars (which are similarly high-order low-degree
g-mode pulsators with roughly half the mass of KIC\,7760680) studied by \cite{kurtz-2014-01} and 
\cite{saio-2015-01} unambiguously exhibit core-to-surface rotation frequency close to unity, from the
frequency splittings of their p- and g-modes.
Star\,\textsc{I} was also shown to be a very slowly rotating star with a counter rotating envelope with 
respect to its core \citep{triana-2015-01}. 
\cite{rogers-2015-01} successfully explained all these observed cases through very efficient angular 
momentum transport carried by internal gravity waves \citep[see also][]{zahn-1997-01,talon-2005-01,
rogers-2013-01}.
The same mechanism is used by \cite{aerts-2015-02} to explain the observed background power excess in the
periodograms of three CoRoT O-type dwarfs.
Therefore, in dwarf stars earlier than A-type, the angular momentum transport can be more efficient than 
predicted, and it can impose near-rigid envelope rotation.

The two recently modelled {\it Kepler} SPB stars, Star\,\textsc{I} and KIC\,7760680 are the best understood stars
of their class.
Their constrained physical parameters can serve as starting point for more sophisticated and/or realistic 
theories of energy and chemical transport by turbulent convection, beyond MLT.
In fact, more realistic future convection models should succeed to improve fitting the observed frequencies of 
these two SPB stars, in addition to surviving the helioseismic tests.

%%%%%%%%%%%%%%%%%%%%%%%%%%%%%%%%%%%%%%%%%%%%%%%%%%%%%%%%%%%%%%%%%%%%%%%%%%%%%%%%%%%%%%%%%%%%%%%%%%%%
\acknowledgments
The authors are grateful to Peter I. P\'apics for the reanalysis of the light curve, and to determine the 
correction factor to be applied for frequency uncertainties ($Q=4.0$) compared to those derived from NLLS.
The research leading to these results has received funding from the People Programme (Marie
Curie Actions) of the European Union's Seventh Framework Programme FP7/2007-2013/ under REA
grant agreement N$^\circ$\,623303 (ASAMBA), from the European
Research Council (ERC) under the European Union's Horizon 2020 research and
innovation programme grant N$^\circ$\,670519 (MAMSIE), and from the European
Community's Seventh Framework Programme FP7-SPACE-2011-1, project N$^\circ$\,312844
(SPACEINN).
RHDT acknowledges support from NSF under the SI$^{2}$ program grants (ACI-1339600) and NASA under the 
TCAN program grants (NNX14AB55G).
SM acknowledges funding by the ERC grant N$^\circ$\,647383 (SPIRE).
The computational resources and services used in this work were provided by the 
VSC (Flemish Supercomputer Center), funded by the Hercules Foundation and the Flemish
Government - department EWI.

%%%%%%%%%%%%%%%%%%%%%%%%%%%%%%%%%%%%%%%%%%%%%%%%%%%%%%%%%%%%%%%%%%%%%%%%%%%%%%%%%%%%%%%%%%%%%%%%%%%%

\appendix

%%%%%%%%%%%%%%%%%%%%%%%%%%%%%%%%%%%%%%%%%%%%%%%%%%%%%%%%%%%%%%%%%%%%%%%%%%%%%%%%%%%%%%%%%%%%%%%%%%%%

\section{Distinguishing Harmonic Degree and Azimuthal Order}\label{ap-em}

\begin{figure}[t!]
\includegraphics[width=\columnwidth]{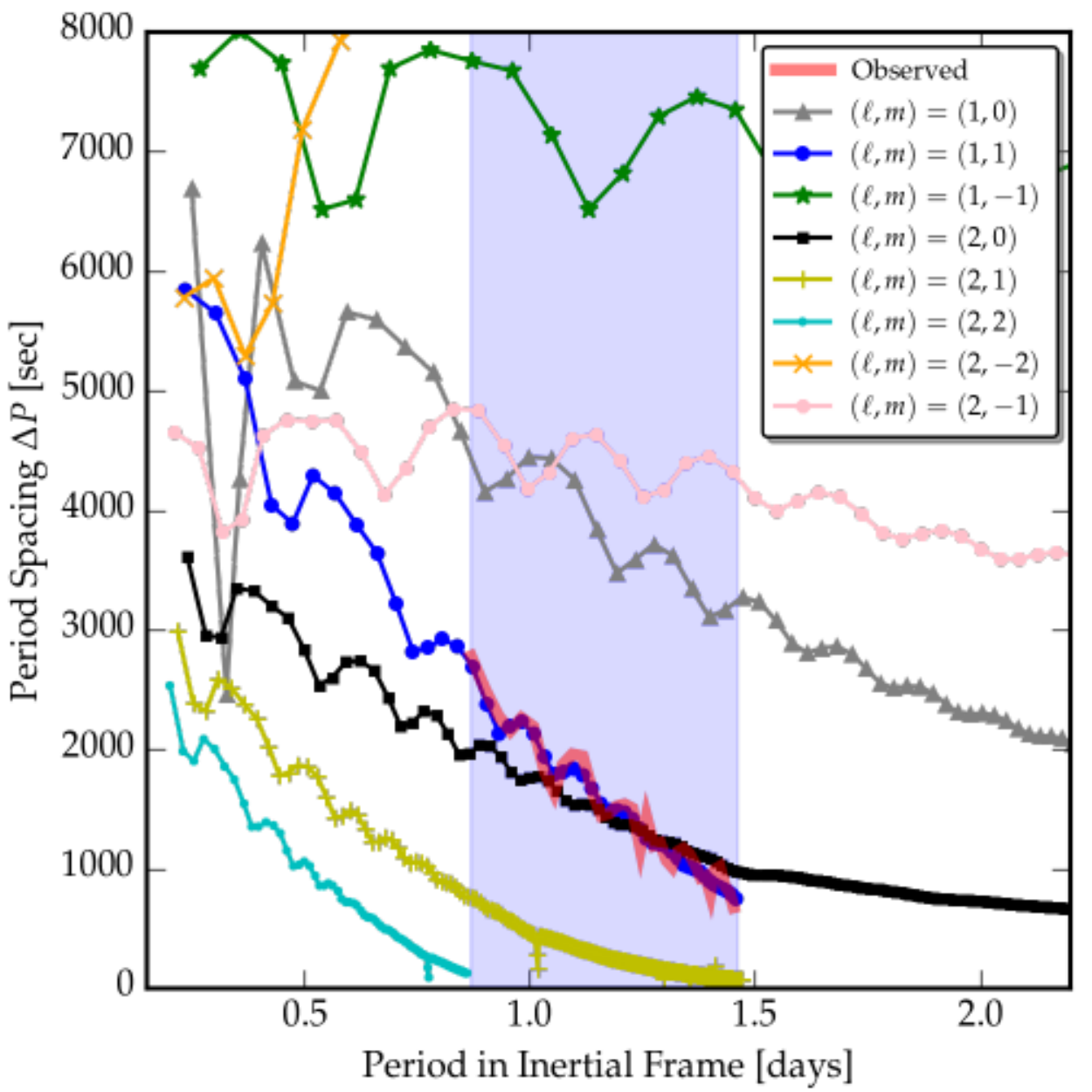} 
\caption{Period spacing $\Delta P$ for eight combinations of $(\ell, \,m)$ for dipole and quadrupole
modes, compared with the observed series (red solid line).
The input model is set to $\sim24\%$ rotation frequency with respect to $f_{\rm rot}^{\rm (Rch)}$.
\label{f-dP-all-l-m}}
\end{figure}

Fig.\,\ref{f-dP-all-l-m} shows eight period spacing series for dipole and quadrupole modes using TAR.
The input model fulfils the position of the star on the Kiel diagram, and is set to uniform rotation
with 24.2\% with respect to Roche critical frequency (Eq.\,\ref{e-f-crit}).
The observed $\Delta P$ pattern (red line) is well reproduced by dipole prograde modes, while all other 
spacings fail to satisfy Eq.\,\ref{e-frot-opt}, and match the slope of the period spacing.
Moreover, extremely high radial orders of up to $|n_{\rm pg}\approx700|$ were needed for 
$(\ell, \,m)=(2, \,1)$ and $(2, \,2)$ to force them towards the observed range, which contradicts
with the requirement $\mathcal{N}=36$.
Thus, the observed serie in Fig.\,\ref{f-obs-dP}a is identified as dipole prograde $(\ell, \,m)=(1, \,+1)$ 
g-modes.

% Figure %
% Figure %

%%%%%%%%%%%%%%%%%%%%%%%%%%%%%%%%%%%%%%%%%%%%%%%%%%%%%%%%%%%%%%%%%%%%%%%%%%%%%%%%%%%%%%%%%%%%%%%%%%%%

\section{Deliverables, inlists and opacity tables}\label{ap-files}

Following the MESA users' code of conduct stated in \cite{paxton-2011-01}, we publish the MESA and GYRE
inlists, and the structure file of the best asteroseismic model of KIC\,7760680 (compatible with GYRE).
This ensures the reproducibility of our results, provided similar MESA and GYRE versions are used.
These products are availble through the following static link:
\href{https://fys.kuleuven.be/ster/Projects/ASAMBA}{https://fys.kuleuven.be/ster/Projects/ASAMBA}.
The adopted OP opacity tables are already available from 
\href{https://bitbucket.org/ehsan_moravveji/op_mono}{https://bitbucket.org/ehsan\_moravveji/op\_mono}.

%%%%%%%%%%%%%%%%%%%%%%%%%%%%%%%%%%%%%%%%%%%%%%%%%%%%%%%%%%%%%%%%%%%%%%%%%%%%%%%%%%%%%%%%%%%%%%%%%%%%

\section{Iterative Procedure to Optimize $f_{\rm rot}$}\label{ap-optim}

\begin{figure}[th!]
\includegraphics[width=\columnwidth]{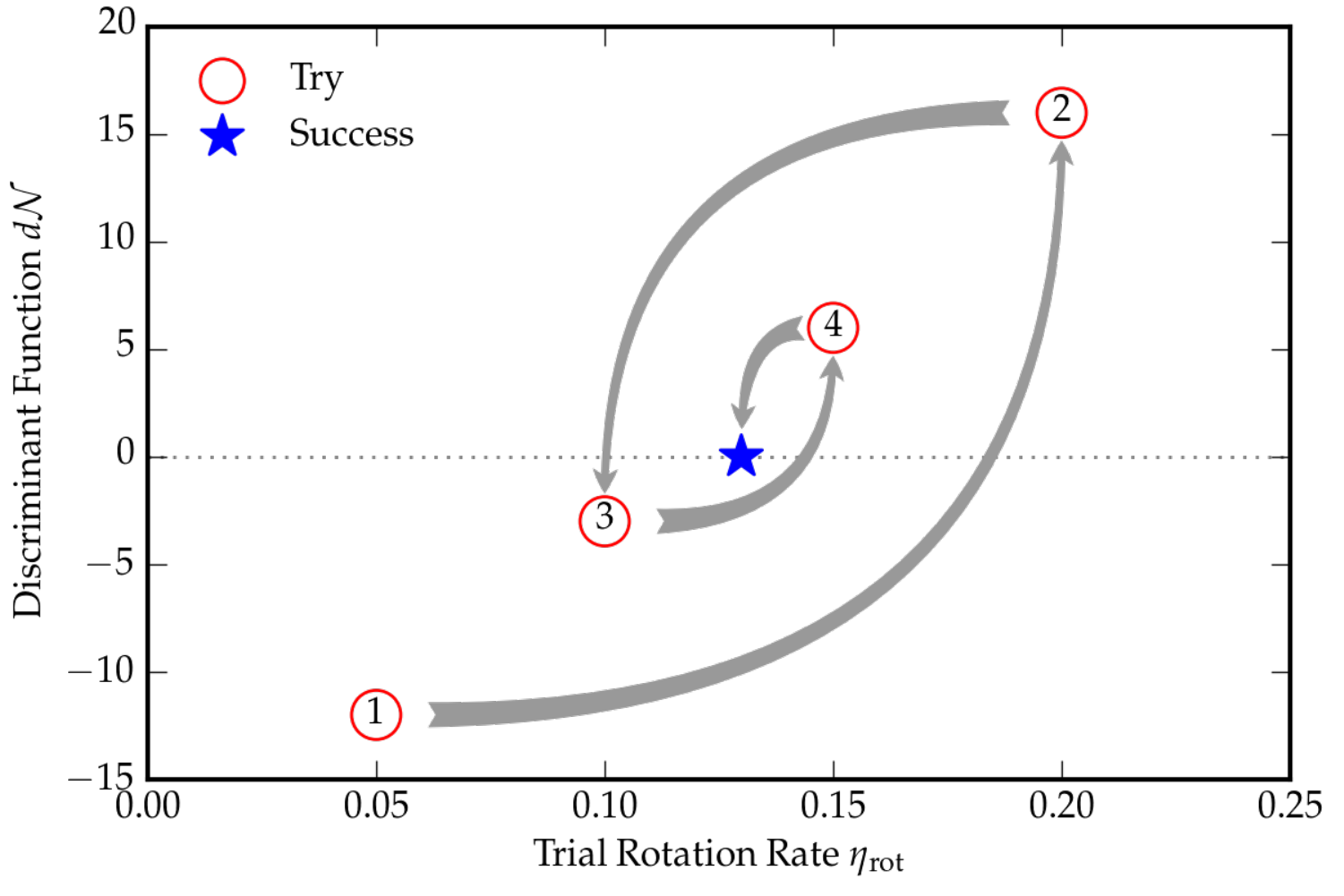} 
\caption{The scheme for optimizing $\eta_{\rm rot}=f_{\rm rot}/f_{\rm rot}^{\rm(Rch)}$. 
\label{f-optim}}
\end{figure}

Here, we explain the iterative procedure to optimise $f_{\rm rot}$ (or equivalently $\eta_{\rm rot}$) 
using Eqs.\,(\ref{e-dN}) and (\ref{e-frot-opt}).
Fig.\,\ref{f-optim} illustrates the procedure.
The first attempt corresponds to a small trial $f_{\rm rot}$, and the second one corresponds to
a much larger trial value for $f_{\rm rot}$, ensuring a change of sign of $d\mathcal{N}$.
For the third attempt, we estimate $f_{\rm rot}$ by Newton-Raphson root-finding, assuming a line
connecting the first two points. 
From the fourth attempt onward, we use the Van Wijngaarden-Dekker-Brent \citep{brent-1973-book} 
root-finding algorithm to locate the zero of $d\mathcal{N}$ \citep[see also][]{press-2007-book}.
The iterations proceed, until the root is successfully located.
Considering the fact that $d\mathcal{N}$ is an integer-valued function, it seldom happens that a tiny 
change in $f_{\rm rot}$ proposed by the previous (Brent) guess does not change 
$\mathcal{M}$.
In such cases $d\mathcal{N}$ is zero (has a staircase shape), and the Brent scheme diverges.
To avoid such circumstances, we employ an iterative Bisection method \citep{press-2007-book} to converge 
to the root of $d\mathcal{N}$.
During all these attempts, we call GYRE, and store the intermediate results on the disk, until the 
procedure succeeds.
Finally, we store the optimal rotation frequency $f_{\rm rot}^{\rm(opt)}$ as an additional attribute 
in the GYRE output summary file.

%%%%%%%%%%%%%%%%%%%%%%%%%%%%%%%%%%%%%%%%%%%%%%%%%%%%%%%%%%%%%%%%%%%%%%%%%%%%%%%%%%%%%%%%%%%%%%%%%%%%
% BIBLIOGRAPHY 
\bibliographystyle{apj}
\bibliography{bib.bib}

\begin{thebibliography}{}
\expandafter\ifx\csname natexlab\endcsname\relax\def\natexlab#1{#1}\fi

\bibitem[{{Aerts}(2013)}]{aerts-2013-01}
{Aerts}, C. 2013, in EAS Publications Series, Vol.~64, EAS Publications Series,
  323--330

\bibitem[{{Aerts} {et~al.}(2010){Aerts}, {Christensen-Dalsgaard}, \&
  {Kurtz}}]{aerts-2010-book}
{Aerts}, C., {Christensen-Dalsgaard}, J., \& {Kurtz}, D.~W. 2010,
  {Asteroseismology, Astronomy and Astrophsyics Library, Springer Berlin
  Heidelberg}

\bibitem[{{Aerts} {et~al.}(2004){Aerts}, {Lamers}, \&
  {Molenberghs}}]{aerts-2004-02}
{Aerts}, C., {Lamers}, H.~J.~G.~L.~M., \& {Molenberghs}, G. 2004, \aap, 418,
  639

\bibitem[{{Aerts} \& {Rogers}(2015)}]{aerts-2015-02}
{Aerts}, C., \& {Rogers}, T.~M. 2015, \apjl, 806, L33

\bibitem[{{Aprilia} {et~al.}(2011){Aprilia}, {Lee}, \&
  {Saio}}]{aprilia-2011-01}
{Aprilia}, {Lee}, U., \& {Saio}, H. 2011, \mnras, 412, 2265

\bibitem[{{Arnett}(2014)}]{arnett-2014-01}
{Arnett}, W.~D. 2014, in {New Windows on Massive Stars: Asteroseismology,
  Interferometry and Spectropolarimetry}, ed. G.~{Meynet}, C.~{Georgy},
  J.~{Groh}, \& P.~{Stee} (Cambridge University Press)

\bibitem[{{Arnett} {et~al.}(2015){Arnett}, {Meakin}, {Viallet}, {Campbell},
  {Lattanzio}, \& {Moc{\'a}k}}]{arnett-2015-01}
{Arnett}, W.~D., {Meakin}, C., {Viallet}, M., {et~al.} 2015, The Astrophysical
  Journal, 809, 30

\bibitem[{{Asplund} {et~al.}(2009){Asplund}, {Grevesse}, {Sauval}, \&
  {Scott}}]{asplund-2009-01}
{Asplund}, M., {Grevesse}, N., {Sauval}, A.~J., \& {Scott}, P. 2009, \araa, 47,
  481

\bibitem[{{Ausseloos} {et~al.}(2004){Ausseloos}, {Scuflaire}, {Thoul}, \&
  {Aerts}}]{ausseloos-2004-01}
{Ausseloos}, M., {Scuflaire}, R., {Thoul}, A., \& {Aerts}, C. 2004, \mnras,
  355, 352

\bibitem[{{Badnell} {et~al.}(2005){Badnell}, {Bautista}, {Butler}, {Delahaye},
  {Mendoza}, {Palmeri}, {Zeippen}, \& {Seaton}}]{badnell-2005-01}
{Badnell}, N.~R., {Bautista}, M.~A., {Butler}, K., {et~al.} 2005, \mnras, 360,
  458

\bibitem[{{Bailey} {et~al.}(2015){Bailey}, {Nagayama}, {Loisel}, {Rochau},
  {Blancard}, {Colgan}, {Cosse}, {Faussurier}, {Fontes}, {Gilleron},
  {Golvokin}, {Hansen}, {Iglesias}, {Kilcrease}, {MacFarlane}, {Mancini},
  {Nahar}, {Orban}, {Pain}, {Pradhan}, {Sherrill}, \&
  {Wilson}}]{bailey-2015-01}
{Bailey}, J.~E., {Nagayama}, T., {Loisel}, G.~P., {et~al.} 2015, \nat, 517, 56

\bibitem[{{Ballot} {et~al.}(2012){Ballot}, {Ligni{\`e}res}, {Prat}, {Reese}, \&
  {Rieutord}}]{ballot-2012-01}
{Ballot}, J., {Ligni{\`e}res}, F., {Prat}, V., {Reese}, D.~R., \& {Rieutord},
  M. 2012, in Astronomical Society of the Pacific Conference Series, Vol. 462,
  Progress in Solar/Stellar Physics with Helio- and Asteroseismology, ed.
  H.~{Shibahashi}, M.~{Takata}, \& A.~E. {Lynas-Gray}, 389

\bibitem[{{Ballot} {et~al.}(2010){Ballot}, {Ligni{\`e}res}, {Reese}, \&
  {Rieutord}}]{ballot-2010-01}
{Ballot}, J., {Ligni{\`e}res}, F., {Reese}, D.~R., \& {Rieutord}, M. 2010,
  \aap, 518, A30

\bibitem[{{Balona} {et~al.}(2011){Balona}, {Pigulski}, {Cat}, {Handler},
  {Guti{\'e}rrez-Soto}, {Engelbrecht}, {Frescura}, {Briquet}, {Cuypers},
  {Daszy{\'n}ska-Daszkiewicz}, {Degroote}, {Dukes}, {Garcia}, {Green}, {Heber},
  {Kawaler}, {Lehmann}, {Leroy}, {Molenda-{\.Z}aaowicz}, {Neiner}, {Noels},
  {Nuspl}, {{\O}stensen}, {Pricopi}, {Roxburgh}, {Salmon}, {Smith},
  {Su{\'a}rez}, {Suran}, {Szab{\'o}}, {Uytterhoeven}, {Christensen-Dalsgaard},
  {Kjeldsen}, {Caldwell}, {Girouard}, \& {Sanderfer}}]{balona-2011-01}
{Balona}, L.~A., {Pigulski}, A., {Cat}, P.~D., {et~al.} 2011, \mnras, 413, 2403

\bibitem[{{Basu}(2014)}]{basu-2014-01}
{Basu}, S. 2014, {Studying stars through frequency inversions}, ed. P.~L.
  {Pall{\'e}} \& C.~{Esteban}, 87

\bibitem[{{Belyaev} {et~al.}(2015){Belyaev}, {Quataert}, \&
  {Fuller}}]{belyaev-2015-01}
{Belyaev}, M.~A., {Quataert}, E., \& {Fuller}, J. 2015, \mnras, 452, 2700

\bibitem[{{B{\"o}hm-Vitense}(1958)}]{bohm-vitense-1958-01}
{B{\"o}hm-Vitense}, E. 1958, \zap, 46, 108

\bibitem[{{Bouabid} {et~al.}(2013){Bouabid}, {Dupret}, {Salmon},
  {Montalb{\'a}n}, {Miglio}, \& {Noels}}]{bouabid-2013-01}
{Bouabid}, M.-P., {Dupret}, M.-A., {Salmon}, S., {et~al.} 2013, \mnras, 429,
  2500

\bibitem[{Brent(1973)}]{brent-1973-book}
Brent, R.~P. 1973, Algorithms for minimization without derivatives,
  Prentice-Hall series in automatic computation (Englewood Cliffs, N.J.
  Prentice-Hall)

\bibitem[{{Briquet} {et~al.}(2007){Briquet}, {Morel}, {Thoul}, {Scuflaire},
  {Miglio}, {Montalb{\'a}n}, {Dupret}, \& {Aerts}}]{briquet-2007-01}
{Briquet}, M., {Morel}, T., {Thoul}, A., {et~al.} 2007, \mnras, 381, 1482

\bibitem[{{Briquet} {et~al.}(2016){Briquet}, {Neiner}, {Petit}, {Leroy}, {de
  Batz}, \& {the MiMeS collaboration}}]{briquet-2016-01}
{Briquet}, M., {Neiner}, C., {Petit}, P., {et~al.} 2016, \aap, 587, A126

\bibitem[{{Briquet} {et~al.}(2012){Briquet}, {Neiner}, {Aerts}, {Morel},
  {Mathis}, {Reese}, {Lehmann}, {Costero}, {Echevarria}, {Handler}, {Kambe},
  {Hirata}, {Masuda}, {Wright}, {Yang}, {Pintado}, {Mkrtichian}, {Lee}, {Han},
  {Bruch}, {De Cat}, {Uytterhoeven}, {Lefever}, {Vanautgaerden}, {de Batz},
  {Fr{\'e}mat}, {Henrichs}, {Geers}, {Martayan}, {Hubert}, {Thizy}, \&
  {Tijani}}]{briquet-2012-01}
{Briquet}, M., {Neiner}, C., {Aerts}, C., {et~al.} 2012, \mnras, 427, 483

\bibitem[{{Browning} {et~al.}(2004){Browning}, {Brun}, \&
  {Toomre}}]{browning-2004-01}
{Browning}, M.~K., {Brun}, A.~S., \& {Toomre}, J. 2004, \apj, 601, 512

\bibitem[{{Buldgen} {et~al.}(2015){Buldgen}, {Reese}, \&
  {Dupret}}]{buldgen-2015-02}
{Buldgen}, G., {Reese}, D.~R., \& {Dupret}, M.~A. 2015, \aap, 583, A62

\bibitem[{{Canuto}(2011{\natexlab{a}})}]{canuto-2011-01}
{Canuto}, V.~M. 2011{\natexlab{a}}, \aap, 528, A76

\bibitem[{{Canuto}(2011{\natexlab{b}})}]{canuto-2011-05}
---. 2011{\natexlab{b}}, \aap, 528, A80

\bibitem[{{Castelli} \& {Kurucz}(2003)}]{castelli-2003-01}
{Castelli}, F., \& {Kurucz}, R.~L. 2003, in IAU Symposium, Vol. 210, Modelling
  of Stellar Atmospheres, ed. N.~{Piskunov}, W.~W. {Weiss}, \& D.~F. {Gray},
  A20

\bibitem[{{Cox} \& {Giuli}(1968)}]{cox-1968-01}
{Cox}, J.~P., \& {Giuli}, R.~T. 1968, {Principles of stellar structure}, ed.
  {Cox, J.~P.~\& Giuli, R.~T.}

\bibitem[{{Cunha} {et~al.}(2015){Cunha}, {Stello}, {Avelino},
  {Christensen-Dalsgaard}, \& {Townsend}}]{cunha-2015-01}
{Cunha}, M.~S., {Stello}, D., {Avelino}, P.~P., {Christensen-Dalsgaard}, J., \&
  {Townsend}, R.~H.~D. 2015, \apj, 805, 127

\bibitem[{{De Cat} \& {Aerts}(2002)}]{de-cat-2002-01}
{De Cat}, P., \& {Aerts}, C. 2002, \aap, 393, 965

\bibitem[{{Decressin} {et~al.}(2009){Decressin}, {Mathis}, {Palacios}, {Siess},
  {Talon}, {Charbonnel}, \& {Zahn}}]{decressin-2009-01}
{Decressin}, T., {Mathis}, S., {Palacios}, A., {et~al.} 2009, \aap, 495, 271

\bibitem[{{Degroote} {et~al.}(2009){Degroote}, {Briquet}, {Catala},
  {Uytterhoeven}, {Lefever}, {Morel}, {Aerts}, {Carrier}, {Auvergne}, {Baglin},
  \& {Michel}}]{degroote-2009-01}
{Degroote}, P., {Briquet}, M., {Catala}, C., {et~al.} 2009, \aap, 506, 111

\bibitem[{{Deheuvels} {et~al.}(2016){Deheuvels}, {Brand{\~a}o}, {Silva
  Aguirre}, {Ballot}, {Michel}, {Cunha}, {Lebreton}, \&
  {Appourchaux}}]{deheuvels-2016-01}
{Deheuvels}, S., {Brand{\~a}o}, I., {Silva Aguirre}, V., {et~al.} 2016,
  arXiv:1603.02332

\bibitem[{{Dintrans} \& {Rieutord}(2000)}]{dintrans-2000-01}
{Dintrans}, B., \& {Rieutord}, M. 2000, \aap, 354, 86

\bibitem[{{Dufton} {et~al.}(2013){Dufton}, {Langer}, {Dunstall}, {Evans},
  {Brott}, {de Mink}, {Howarth}, {Kennedy}, {McEvoy}, {Potter},
  {Ram{\'{\i}}rez-Agudelo}, {Sana}, {Sim{\'o}n-D{\'{\i}}az}, {Taylor}, \&
  {Vink}}]{dufton-2013-01}
{Dufton}, P.~L., {Langer}, N., {Dunstall}, P.~R., {et~al.} 2013, \aap, 550,
  A109

\bibitem[{{Dziembowski} \& {Goode}(1992)}]{dziembowski-1992-01}
{Dziembowski}, W.~A., \& {Goode}, P.~R. 1992, \apj, 394, 670

\bibitem[{{Dziembowski} {et~al.}(1993){Dziembowski}, {Moskalik}, \&
  {Pamyatnykh}}]{dziembowski-1993-02}
{Dziembowski}, W.~A., {Moskalik}, P., \& {Pamyatnykh}, A.~A. 1993, \mnras, 265,
  588

\bibitem[{{Dziembowski} \& {Pamyatnykh}(1991)}]{dziembowski-1991-01}
{Dziembowski}, W.~A., \& {Pamyatnykh}, A.~A. 1991, \aap, 248, L11

\bibitem[{{Dziembowski} \& {Pamyatnykh}(2008)}]{dziembowski-2008-01}
---. 2008, \mnras, 385, 2061

\bibitem[{{Eckart}(1960)}]{eckart-1960-book}
{Eckart}, C. 1960, {Hydrodynamics of Oceans and Atmospheres, Pergamon Press,
  Oxford}

\bibitem[{{Endal} \& {Sofia}(1976)}]{endal-1976-01}
{Endal}, A.~S., \& {Sofia}, S. 1976, \apj, 210, 184

\bibitem[{{Endal} \& {Sofia}(1978)}]{endal-1978-01}
---. 1978, \apj, 220, 279

\bibitem[{{Espinosa Lara} \& {Rieutord}(2013)}]{espinosa-lara-2013-01}
{Espinosa Lara}, F., \& {Rieutord}, M. 2013, \aap, 552, A35

\bibitem[{{Freytag} {et~al.}(1996){Freytag}, {Ludwig}, \&
  {Steffen}}]{freytag-1996-01}
{Freytag}, B., {Ludwig}, H.-G., \& {Steffen}, M. 1996, \aap, 313, 497

\bibitem[{{Fuller} {et~al.}(2015){Fuller}, {Cantiello}, {Stello}, {Garcia}, \&
  {Bildsten}}]{fuller-2015-01}
{Fuller}, J., {Cantiello}, M., {Stello}, D., {Garcia}, R.~A., \& {Bildsten}, L.
  2015, Science, 350, 423

\bibitem[{{Gautschy} \& {Saio}(1993)}]{gautschy-1993-01}
{Gautschy}, A., \& {Saio}, H. 1993, \mnras, 262, 213

\bibitem[{{Heger} {et~al.}(2000){Heger}, {Langer}, \&
  {Woosley}}]{heger-2000-01}
{Heger}, A., {Langer}, N., \& {Woosley}, S.~E. 2000, \apj, 528, 368

\bibitem[{{Heger} {et~al.}(2005){Heger}, {Woosley}, \&
  {Spruit}}]{heger-2005-01}
{Heger}, A., {Woosley}, S.~E., \& {Spruit}, H.~C. 2005, \apj, 626, 350

\bibitem[{{Herwig}(2000)}]{herwig-2000-01}
{Herwig}, F. 2000, \aap, 360, 952

\bibitem[{{Huang} {et~al.}(2010){Huang}, {Gies}, \& {McSwain}}]{huang-2010-01}
{Huang}, W., {Gies}, D.~R., \& {McSwain}, M.~V. 2010, \apj, 722, 605

\bibitem[{{Huat} {et~al.}(2009){Huat}, {Hubert}, {Baudin}, {Floquet}, {Neiner},
  {Fr{\'e}mat}, {Guti{\'e}rrez-Soto}, {Andrade}, {de Batz}, {Diago}, {Emilio},
  {Espinosa Lara}, {Fabregat}, {Janot-Pacheco}, {Leroy}, {Martayan}, {Semaan},
  {Suso}, {Auvergne}, {Catala}, {Michel}, \& {Samadi}}]{huat-2009-01}
{Huat}, A.-L., {Hubert}, A.-M., {Baudin}, F., {et~al.} 2009, \aap, 506, 95

\bibitem[{{Kippenhahn} \& {Thomas}(1970)}]{kippenhahn-1970-01}
{Kippenhahn}, R., \& {Thomas}, H.-C. 1970, in IAU Colloq. 4: Stellar Rotation,
  ed. A.~{Slettebak}, 20

\bibitem[{{Kurtz} {et~al.}(2014){Kurtz}, {Saio}, {Takata}, {Shibahashi},
  {Murphy}, \& {Sekii}}]{kurtz-2014-01}
{Kurtz}, D.~W., {Saio}, H., {Takata}, M., {et~al.} 2014, \mnras, 444, 102

\bibitem[{{Langer} {et~al.}(1985){Langer}, {El Eid}, \&
  {Fricke}}]{langer-1985-01}
{Langer}, N., {El Eid}, M.~F., \& {Fricke}, K.~J. 1985, \aap, 145, 179

\bibitem[{{Ledoux}(1951)}]{ledoux-1951-01}
{Ledoux}, P. 1951, \apj, 114, 373

\bibitem[{{Lee} {et~al.}(2016){Lee}, {Mathis}, \& {Neiner}}]{lee-2016-01}
{Lee}, U., {Mathis}, S., \& {Neiner}, C. 2016, \mnras, 457, 2445

\bibitem[{{Lee} {et~al.}(2014){Lee}, {Neiner}, \& {Mathis}}]{lee-2014-01}
{Lee}, U., {Neiner}, C., \& {Mathis}, S. 2014, \mnras, 443, 1515

\bibitem[{{Lee} \& {Saio}(1986)}]{lee-1986-01}
{Lee}, U., \& {Saio}, H. 1986, \mnras, 221, 365

\bibitem[{{Lee} \& {Saio}(1993)}]{lee-1993-01}
---. 1993, \mnras, 261, 415

\bibitem[{Lee \& Saio(1997)}]{lee-1997-01}
Lee, U., \& Saio, H. 1997, The Astrophysical Journal, 491, 839

\bibitem[{{Maeder}(1975)}]{maeder-1975-01}
{Maeder}, A. 1975, \aap, 40, 303

\bibitem[{{Maeder}(2009)}]{maeder-2009-book}
---. 2009, {Physics, Formation and Evolution of Rotating Stars}, ed. {Maeder,
  A.}, doi:10.1007/978-3-540-76949-1

\bibitem[{{Maeder} {et~al.}(2013){Maeder}, {Meynet}, {Lagarde}, \&
  {Charbonnel}}]{maeder-2013-01}
{Maeder}, A., {Meynet}, G., {Lagarde}, N., \& {Charbonnel}, C. 2013, \aap, 553,
  A1

\bibitem[{{Maeder} \& {Zahn}(1998)}]{maeder-1998-01}
{Maeder}, A., \& {Zahn}, J.-P. 1998, \aap, 334, 1000

\bibitem[{Mathis(2013)}]{mathis-2013-01}
Mathis, S. 2013, Studying Stellar Rotation and Convection: Theoretical
  Background and Seismic Diagnostics, ed. M.~Goupil, K.~Belkacem, C.~Neiner,
  F.~Ligni{\`e}res, \& J.~J. Green (Berlin, Heidelberg: Springer Berlin
  Heidelberg), 23--47

\bibitem[{{Mathis} {et~al.}(2014){Mathis}, {Neiner}, \& {Tran
  Minh}}]{mathis-2014-01}
{Mathis}, S., {Neiner}, C., \& {Tran Minh}, N. 2014, \aap, 565, A47

\bibitem[{{Mathis} {et~al.}(2004){Mathis}, {Palacios}, \&
  {Zahn}}]{mathis-2004-01}
{Mathis}, S., {Palacios}, A., \& {Zahn}, J.-P. 2004, \aap, 425, 243

\bibitem[{{Mathis} {et~al.}(2008){Mathis}, {Talon}, {Pantillon}, \&
  {Zahn}}]{mathis-2008-01}
{Mathis}, S., {Talon}, S., {Pantillon}, F.-P., \& {Zahn}, J.-P. 2008, \solphys,
  251, 101

\bibitem[{{Mathis} \& {Zahn}(2004)}]{mathis-2004-02}
{Mathis}, S., \& {Zahn}, J.-P. 2004, \aap, 425, 229

\bibitem[{{Mathis} \& {Zahn}(2005)}]{mathis-2005-01}
---. 2005, \aap, 440, 653

\bibitem[{{Meynet} \& {Maeder}(2000)}]{meynet-2000-01}
{Meynet}, G., \& {Maeder}, A. 2000, \aap, 361, 101

\bibitem[{{Miglio} {et~al.}(2008){Miglio}, {Montalb{\'a}n}, {Noels}, \&
  {Eggenberger}}]{miglio-2008-01}
{Miglio}, A., {Montalb{\'a}n}, J., {Noels}, A., \& {Eggenberger}, P. 2008,
  \mnras, 386, 1487

\bibitem[{{Mondet} {et~al.}(2015){Mondet}, {Blancard}, {Coss{\'e}}, \&
  {Faussurier}}]{mondet-2015-01}
{Mondet}, G., {Blancard}, C., {Coss{\'e}}, P., \& {Faussurier}, G. 2015, \apjs,
  220, 2

\bibitem[{{Moravveji}(2015)}]{moravveji-2015-02}
{Moravveji}, E. 2015, in EAS Publications Series, Vol.~71, EAS Publications
  Series, 317--320

\bibitem[{{Moravveji}(2016)}]{moravveji-2016-01}
{Moravveji}, E. 2016, \mnras, 455, L67

\bibitem[{{Moravveji} {et~al.}(2015){Moravveji}, {Aerts}, {P\'apics}, {Triana},
  \& {Vandoren}}]{moravveji-2015-01}
{Moravveji}, E., {Aerts}, C., {P\'apics}, P.~I., {Triana}, S.~A., \&
  {Vandoren}, B. 2015, \aap, 580, A27

\bibitem[{{Nagayama} {et~al.}(2016){Nagayama}, {Bailey}, {Loisel}, {Rochau},
  {MacFarlane}, \& {Golovkin}}]{nagayama-2016-01}
{Nagayama}, T., {Bailey}, J.~E., {Loisel}, G., {et~al.} 2016, \pre, 93, 023202

\bibitem[{{Neiner} {et~al.}(2012){Neiner}, {Floquet}, {Samadi}, {Espinosa
  Lara}, {Fr{\'e}mat}, {Mathis}, {Leroy}, {de Batz}, {Rainer}, {Poretti},
  {Mathias}, {Guarro Fl{\'o}}, {Buil}, {Ribeiro}, {Alecian}, {Andrade},
  {Briquet}, {Diago}, {Emilio}, {Fabregat}, {Guti{\'e}rrez-Soto}, {Hubert},
  {Janot-Pacheco}, {Martayan}, {Semaan}, {Suso}, \& {Zorec}}]{neiner-2012-01}
{Neiner}, C., {Floquet}, M., {Samadi}, R., {et~al.} 2012, \aap, 546, A47

\bibitem[{{Nieva} \& {Przybilla}(2012)}]{nieva-2012-01}
{Nieva}, M.-F., \& {Przybilla}, N. 2012, \aap, 539, A143

\bibitem[{{Pamyatnykh}(1999)}]{pamyatnykh-1999-01}
{Pamyatnykh}, A.~A. 1999, \actaa, 49, 119

\bibitem[{{Pantillon} {et~al.}(2007){Pantillon}, {Talon}, \&
  {Charbonnel}}]{pantillon-2007-01}
{Pantillon}, F.~P., {Talon}, S., \& {Charbonnel}, C. 2007, \aap, 474, 155

\bibitem[{{P{\'a}pics} {et~al.}(2014){P{\'a}pics}, {Moravveji}, {Aerts},
  {Tkachenko}, {Triana}, {Bloemen}, \& {Southworth}}]{papics-2014-01}
{P{\'a}pics}, P.~I., {Moravveji}, E., {Aerts}, C., {et~al.} 2014, \aap, 570, A8

\bibitem[{{P{\'a}pics} {et~al.}(2015){P{\'a}pics}, {Tkachenko}, {Aerts}, {Van
  Reeth}, {De Smedt}, {Hillen}, {{\O}stensen}, \& {Moravveji}}]{papics-2015-01}
{P{\'a}pics}, P.~I., {Tkachenko}, A., {Aerts}, C., {et~al.} 2015, \apjl, 803,
  L25

\bibitem[{{P{\'a}pics} {et~al.}(2011){P{\'a}pics}, {Briquet}, {Auvergne},
  {Aerts}, {Degroote}, {Niemczura}, {Vu{\v c}kovi{\'c}}, {Smolders}, {Poretti},
  {Rainer}, {Hareter}, {Baglin}, {Baudin}, {Catala}, {Michel}, \&
  {Samadi}}]{papics-2011-01}
{P{\'a}pics}, P.~I., {Briquet}, M., {Auvergne}, M., {et~al.} 2011, \aap, 528,
  A123+

\bibitem[{{P{\'a}pics} {et~al.}(2012){P{\'a}pics}, {Briquet}, {Baglin},
  {Poretti}, {Aerts}, {Degroote}, {Tkachenko}, {Morel}, {Zima}, {Niemczura},
  {Rainer}, {Hareter}, {Baudin}, {Catala}, {Michel}, {Samadi}, \&
  {Auvergne}}]{papics-2012-01}
{P{\'a}pics}, P.~I., {Briquet}, M., {Baglin}, A., {et~al.} 2012, \aap, 542, A55

\bibitem[{{Pasetto} {et~al.}(2015){Pasetto}, {Chiosi}, {Chiosi}, {Cropper}, \&
  {Weiss}}]{pasetto-2015-01}
{Pasetto}, S., {Chiosi}, C., {Chiosi}, E., {Cropper}, M., \& {Weiss}, A. 2015,
  arXiv:1511.08811

\bibitem[{{Pasetto} {et~al.}(2014){Pasetto}, {Chiosi}, {Cropper}, \&
  {Grebel}}]{pasetto-2014-01}
{Pasetto}, S., {Chiosi}, C., {Cropper}, M., \& {Grebel}, E.~K. 2014, \mnras,
  445, 3592

\bibitem[{{Paxton} {et~al.}(2011){Paxton}, {Bildsten}, {Dotter}, {Herwig},
  {Lesaffre}, \& {Timmes}}]{paxton-2011-01}
{Paxton}, B., {Bildsten}, L., {Dotter}, A., {et~al.} 2011, \apjs, 192, 3

\bibitem[{{Paxton} {et~al.}(2013){Paxton}, {Cantiello}, {Arras}, {Bildsten},
  {Brown}, {Dotter}, {Mankovich}, {Montgomery}, {Stello}, {Timmes}, \&
  {Townsend}}]{paxton-2013-01}
{Paxton}, B., {Cantiello}, M., {Arras}, P., {et~al.} 2013, \apjs, 208, 4

\bibitem[{Paxton {et~al.}(2015)Paxton, Marchant, Schwab, Bauer, Bildsten,
  Cantiello, Dessart, Farmer, Hu, Langer, Townsend, Townsley, \&
  Timmes}]{paxton-2015-01}
Paxton, B., Marchant, P., Schwab, J., {et~al.} 2015, \apjs, 220, 15

\bibitem[{{Prat} {et~al.}(2016){Prat}, {Ligni{\`e}res}, \&
  {Ballot}}]{prat-2016-01}
{Prat}, V., {Ligni{\`e}res}, F., \& {Ballot}, J. 2016, \aap, 587, A110

\bibitem[{Press {et~al.}(2007)Press, Teukolsky, Vetterling, \&
  Flannery}]{press-2007-book}
Press, W.~H., Teukolsky, S.~A., Vetterling, W.~T., \& Flannery, B.~P. 2007,
  Numerical Recipes 3rd Edition: The Art of Scientific Computing, 3rd edn. (New
  York, NY, USA: Cambridge University Press)

\bibitem[{{Puls} {et~al.}(2015){Puls}, {Sundqvist}, \&
  {Markova}}]{puls-2015-01}
{Puls}, J., {Sundqvist}, J.~O., \& {Markova}, N. 2015, in IAU Symposium, Vol.
  307, IAU Symposium, ed. G.~{Meynet}, C.~{Georgy}, J.~{Groh}, \& P.~{Stee},
  25--36

\bibitem[{{Ram{\'{\i}}rez-Agudelo} {et~al.}(2013){Ram{\'{\i}}rez-Agudelo},
  {Sim{\'o}n-D{\'{\i}}az}, {Sana}, {de Koter}, {Sab{\'{\i}}n-Sanjul{\'{\i}}an},
  {de Mink}, {Dufton}, {Gr{\"a}fener}, {Evans}, {Herrero}, {Langer}, {Lennon},
  {Ma{\'{\i}}z Apell{\'a}niz}, {Markova}, {Najarro}, {Puls}, {Taylor}, \&
  {Vink}}]{ramirez-2013-01}
{Ram{\'{\i}}rez-Agudelo}, O.~H., {Sim{\'o}n-D{\'{\i}}az}, S., {Sana}, H.,
  {et~al.} 2013, \aap, 560, A29

\bibitem[{{Ram{\'{\i}}rez-Agudelo} {et~al.}(2015){Ram{\'{\i}}rez-Agudelo},
  {Sana}, {de Mink}, {H{\'e}nault-Brunet}, {de Koter}, {Langer}, {Tramper},
  {Gr{\"a}fener}, {Evans}, {Vink}, {Dufton}, \& {Taylor}}]{ramirez-2015-01}
{Ram{\'{\i}}rez-Agudelo}, O.~H., {Sana}, H., {de Mink}, S.~E., {et~al.} 2015,
  \aap, 580, A92

\bibitem[{{Raskin} {et~al.}(2011){Raskin}, {van Winckel}, {Hensberge},
  {Jorissen}, {Lehmann}, {Waelkens}, {Avila}, {de Cuyper}, {Degroote},
  {Dubosson}, {Dumortier}, {Fr{\'e}mat}, {Laux}, {Michaud}, {Morren}, {Perez
  Padilla}, {Pessemier}, {Prins}, {Smolders}, {van Eck}, \&
  {Winkler}}]{raskin-2011-01}
{Raskin}, G., {van Winckel}, H., {Hensberge}, H., {et~al.} 2011, \aap, 526, A69

\bibitem[{{Reese} {et~al.}(2006){Reese}, {Ligni{\`e}res}, \&
  {Rieutord}}]{reese-2006-01}
{Reese}, D., {Ligni{\`e}res}, F., \& {Rieutord}, M. 2006, \aap, 455, 621

\bibitem[{{Rogers}(2015)}]{rogers-2015-01}
{Rogers}, T.~M. 2015, \apjl, 815, L30

\bibitem[{{Rogers} {et~al.}(2013){Rogers}, {Lin}, {McElwaine}, \&
  {Lau}}]{rogers-2013-01}
{Rogers}, T.~M., {Lin}, D.~N.~C., {McElwaine}, J.~N., \& {Lau}, H.~H.~B. 2013,
  \apj, 772, 21

\bibitem[{{Roxburgh}(1965)}]{roxburgh-1965-01}
{Roxburgh}, I.~W. 1965, \mnras, 130, 223

\bibitem[{{Saio} \& {Deupree}(2012)}]{saio-2012-01}
{Saio}, H., \& {Deupree}, R.~G. 2012, in Astronomical Society of the Pacific
  Conference Series, Vol. 462, Progress in Solar/Stellar Physics with Helio-
  and Asteroseismology, ed. H.~{Shibahashi}, M.~{Takata}, \& A.~E.
  {Lynas-Gray}, 398

\bibitem[{{Saio} {et~al.}(2015){Saio}, {Kurtz}, {Takata}, {Shibahashi},
  {Murphy}, {Sekii}, \& {Bedding}}]{saio-2015-01}
{Saio}, H., {Kurtz}, D.~W., {Takata}, M., {et~al.} 2015, \mnras, 447, 3264

\bibitem[{{Salmon} {et~al.}(2012){Salmon}, {Montalb{\'a}n}, {Morel}, {Miglio},
  {Dupret}, \& {Noels}}]{salmon-2012-01}
{Salmon}, S., {Montalb{\'a}n}, J., {Morel}, T., {et~al.} 2012, \mnras, 422,
  3460

\bibitem[{{Savonije}(2005)}]{savonije-2005-01}
{Savonije}, G.~J. 2005, \aap, 443, 557

\bibitem[{{Savonije}(2013)}]{savonije-2013-01}
---. 2013, \aap, 559, A25

\bibitem[{{Schwarzenberg-Czerny}(1991)}]{schwarzenberg-czerny-1991-01}
{Schwarzenberg-Czerny}, A. 1991, \mnras, 253, 198

\bibitem[{{Seaton}(2005)}]{seaton-2005-01}
{Seaton}, M.~J. 2005, \mnras, 362, L1

\bibitem[{{Shibahashi} \& {Ishimatsu}(2013)}]{shibahashi-2013-01}
{Shibahashi}, H., \& {Ishimatsu}, H. 2013, in Astrophysics and Space Science
  Proceedings, Vol.~31, Stellar Pulsations: Impact of New Instrumentation and
  New Insights, ed. J.~C. {Su{\'a}rez}, R.~{Garrido}, L.~A. {Balona}, \&
  J.~{Christensen-Dalsgaard}, 49

\bibitem[{{Soufi} {et~al.}(1998){Soufi}, {Goupil}, \&
  {Dziembowski}}]{soufi-1998-01}
{Soufi}, F., {Goupil}, M.~J., \& {Dziembowski}, W.~A. 1998, \aap, 334, 911

\bibitem[{{Stancliffe} {et~al.}(2015){Stancliffe}, {Fossati}, {Passy}, \&
  {Schneider}}]{stancliffe-2015-01}
{Stancliffe}, R.~J., {Fossati}, L., {Passy}, J.-C., \& {Schneider}, F.~R.~N.
  2015, \aap, 575, A117

\bibitem[{{Stellingwerf}(1978)}]{stellingwerf-1978-02}
{Stellingwerf}, R.~F. 1978, \aj, 83, 1184

\bibitem[{{Stello} {et~al.}(2016){Stello}, {Cantiello}, {Fuller}, {Huber},
  {Garc{\'{\i}}a}, {Bedding}, {Bildsten}, \& {Aguirre}}]{stello-2016-01}
{Stello}, D., {Cantiello}, M., {Fuller}, J., {et~al.} 2016, \nat, 529, 364

\bibitem[{{Talon} \& {Charbonnel}(2005)}]{talon-2005-01}
{Talon}, S., \& {Charbonnel}, C. 2005, \aap, 440, 981

\bibitem[{{Talon} {et~al.}(1997){Talon}, {Zahn}, {Maeder}, \&
  {Meynet}}]{talon-1997-01}
{Talon}, S., {Zahn}, J.-P., {Maeder}, A., \& {Meynet}, G. 1997, \aap, 322, 209

\bibitem[{{Tassoul}(1980)}]{tassoul-1980-01}
{Tassoul}, M. 1980, \apjs, 43, 469

\bibitem[{{Townsend}(2003{\natexlab{a}})}]{townsend-2003-02}
{Townsend}, R.~H.~D. 2003{\natexlab{a}}, \mnras, 343, 125

\bibitem[{{Townsend}(2003{\natexlab{b}})}]{townsend-2003-03}
---. 2003{\natexlab{b}}, \mnras, 340, 1020

\bibitem[{{Townsend}(2005{\natexlab{a}})}]{townsend-2005-01}
---. 2005{\natexlab{a}}, \mnras, 360, 465

\bibitem[{{Townsend}(2005{\natexlab{b}})}]{townsend-2005-02}
---. 2005{\natexlab{b}}, \mnras, 364, 573

\bibitem[{{Townsend} \& {Teitler}(2013)}]{townsend-2013-01}
{Townsend}, R.~H.~D., \& {Teitler}, S.~A. 2013, \mnras, 435, 3406

\bibitem[{{Triana} {et~al.}(2015){Triana}, {Moravveji}, {P\'apics}, {Aerts},
  {Kawaler}, \& {Christensen-Dalsgaard}}]{triana-2015-01}
{Triana}, S.~A., {Moravveji}, E., {P\'apics}, P.~I., {et~al.} 2015, \apj, 810,
  16

\bibitem[{{Unno} {et~al.}(1989){Unno}, {Osaki}, {Ando}, {Saio}, \&
  {Shibahashi}}]{unno-1989-book}
{Unno}, W., {Osaki}, Y., {Ando}, H., {Saio}, H., \& {Shibahashi}, H. 1989,
  {Nonradial oscillations of stars}, ed. {Unno, W., Osaki, Y., Ando, H., Saio,
  H., \& Shibahashi, H.}

\bibitem[{{Van Reeth} {et~al.}(2015{\natexlab{a}}){Van Reeth}, {Tkachenko},
  {Aerts}, {P{\'a}pics}, {Degroote}, {Debosscher}, {Zwintz}, {Bloemen}, {De
  Smedt}, {Hrudkova}, {Raskin}, \& {Van Winckel}}]{vanreeth-2015-01}
{Van Reeth}, T., {Tkachenko}, A., {Aerts}, C., {et~al.} 2015{\natexlab{a}},
  \aap, 574, A17

\bibitem[{{Van Reeth} {et~al.}(2015{\natexlab{b}}){Van Reeth}, {Tkachenko},
  {Aerts}, {P{\'a}pics}, {Triana}, {Zwintz}, {Degroote}, {Debosscher},
  {Bloemen}, {Schmid}, {De Smedt}, {Fremat}, {Fuentes}, {Homan}, {Hrudkova},
  {Karjalainen}, {Lombaert}, {Nemeth}, {{\O}stensen}, {Van De Steene}, {Vos},
  {Raskin}, \& {Van Winckel}}]{vanreeth-2015-02}
---. 2015{\natexlab{b}}, \apjs, 218, 27

\bibitem[{{Viallet} {et~al.}(2015){Viallet}, {Meakin}, {Prat}, \&
  {Arnett}}]{viallet-2015-01}
{Viallet}, M., {Meakin}, C., {Prat}, V., \& {Arnett}, D. 2015, \aap, 580, A61

\bibitem[{{Vink} {et~al.}(2001){Vink}, {de Koter}, \& {Lamers}}]{vink-2001-01}
{Vink}, J.~S., {de Koter}, A., \& {Lamers}, H.~J.~G.~L.~M. 2001, \aap, 369, 574

\bibitem[{{Waelkens}(1991)}]{waelkens-1991-01}
{Waelkens}, C. 1991, \aap, 246, 453

\bibitem[{{Waelkens} {et~al.}(1998){Waelkens}, {Aerts}, {Kestens}, {Grenon}, \&
  {Eyer}}]{waelkens-1998-01}
{Waelkens}, C., {Aerts}, C., {Kestens}, E., {Grenon}, M., \& {Eyer}, L. 1998,
  \aap, 330, 215

\bibitem[{{Xiong}(1979)}]{xiong-1979-01}
{Xiong}, D.-R. 1979, Acta Astronomica Sinica, 20, 238

\bibitem[{{Xiong}(1989)}]{xiong-1989-02}
---. 1989, \aap, 213, 176

\bibitem[{{Zahn}(1991)}]{zahn-1991-01}
{Zahn}, J.-P. 1991, \aap, 252, 179

\bibitem[{{Zahn}(1992)}]{zahn-1992-01}
---. 1992, \aap, 265, 115

\bibitem[{{Zahn} {et~al.}(1997){Zahn}, {Talon}, \& {Matias}}]{zahn-1997-01}
{Zahn}, J.-P., {Talon}, S., \& {Matias}, J. 1997, \aap, 322, 320

\bibitem[{{Zemskova} {et~al.}(2014){Zemskova}, {Garaud}, {Deal}, \&
  {Vauclair}}]{zemskova-2014-01}
{Zemskova}, V., {Garaud}, P., {Deal}, M., \& {Vauclair}, S. 2014, \apj, 795,
  118

\bibitem[{{Zhang}(2013)}]{zhang-2013-01}
{Zhang}, Q.~S. 2013, \apjs, 205, 18

\bibitem[{{Zhang}(2016)}]{zhang-2016-01}
---. 2016, \apj, 818, 146

\bibitem[{{Zhang} \& {Li}(2012{\natexlab{a}})}]{zhang-2012-02}
{Zhang}, Q.~S., \& {Li}, Y. 2012{\natexlab{a}}, \apj, 746, 50

\bibitem[{{Zhang} \& {Li}(2012{\natexlab{b}})}]{zhang-2012-03}
---. 2012{\natexlab{b}}, \apj, 750, 11

\end{thebibliography}
%%%%%%%%%%%%%%%%%%%%%%%%%%%%%%%%%%%%%%%%%%%%%%%%%%%%%%%%%%%%%%%%%%%%%%%%%%%%%%%%%%%%%%%%%%%%%%%%%%%%

\end{document}